\documentclass[%
 aip,
 amsmath,amssymb,
 reprint,%
]{revtex4-1}

\usepackage{ulem}
\usepackage{preamble}

\begin{document}

\preprint{AIP/123-QED}

\title{Double excitation energies from quantum Monte Carlo using state-specific energy optimization}

\author{Stuart Shepard}
\email{s.s.shepard@utwente.nl}
\affiliation{MESA+ Institute for Nanotechnology, University of Twente, 7500 AE Enschede, The Netherlands}

\author{Ram\'{o}n L. Panad\'{e}s-Barrueta}%
\altaffiliation{Current address: Faculty of Chemistry and Food Chemistry, Technische Universit\"{a}t Dresden, 01062 Dresden, Germany}
\affiliation{MESA+ Institute for Nanotechnology, University of Twente, 7500 AE Enschede, The Netherlands}

\author{Saverio Moroni}
\email{moroni@democritos.it}
\affiliation{CNR-IOM DEMOCRITOS, Istituto Officina dei Materiali and SISSA Scuola Internazionale Superiore di Studi Avanzati, Via Bonomea 265, I-34136 Trieste, Italy}

\author{Anthony Scemama}
\email{scemama@irsamc.ups-tlse.fr}
\affiliation{Laboratoire de Chimie et Physique Quantiques, Universit\'{e} de Toulouse, CNRS, UPS, 31062 Toulouse, France}

\author{Claudia Filippi}
\email{c.filippi@utwente.nl}
\affiliation{MESA+ Institute for Nanotechnology, University of Twente, 7500 AE Enschede, The Netherlands}

\date{\today}

\begin{abstract}

We show that recently developed quantum Monte Carlo methods, which provide accurate vertical transition energies for single excitations, also successfully treat double excitations. We study the double excitations in medium-sized molecules, some of which are challenging for high level coupled-cluster calculations to model accurately. Our fixed-node diffusion Monte Carlo excitation energies are in very good agreement with reliable benchmarks, when available, and provide accurate predictions for excitation energies of difficult systems where reference values are lacking. 
\end{abstract}

\maketitle

\section{\label{sec:introduction}Introduction}


An open challenge in computational chemistry is to develop a method which can accurately treat different types of excited states on an equal footing at an affordable cost. An effective way to assess the accuracy of different methods is by comparing vertical transition energies (VTE) with the theoretical best estimates (TBE) determined from high-level computational methods. Environmental effects can influence the measured transition energy in experiment while VTEs calculated using the same molecular geometry are directly comparable.

Conventional single-reference methods, such as time-dependent density functional theory (TD-DFT)~\cite{TDDFT0,TDDFT1,TDDFT2} and coupled-cluster theories (CC) (equation-of-motion/linear response~\cite{EOMCC1}), struggle to model double excitations, i.e., excitations whose determinant expansions are dominated by determinants having two orbitals differing from a given reference determinant.
The most commonly used, best-scaling, version of TD-DFT uses a linear response formalism which cannot treat multi-electron excitations. A relatively uniform description of singly and doubly excited states relies on workarounds and ad-hoc choices of the exchange-correlation functional.~\cite{TDDFT3,TDDFT4,shao_2003,wang_2004,wang_2005,rinkevicius_2010}

In theory, CC can capture double excitations, but in practice the excitation level must be truncated. For an $M$-electron excitation, CC must include at least $M$+1 excitations to include a satisfactory amount of correlation,~\cite{EOMCC2} but often the $M$+2 level theory is needed to obtain reasonable excitation energies.~\cite{questdb,loos2019,questdb2,loos2021,newcc4} CC theories needed to treat single and double excitations 
such as CC3~\cite{CC3} and CCSDT~\cite{CCSDT} (for singles), and CC4~\cite{CC4} and CCSDTQ~\cite{CCSDTQ} (for doubles) scale poorly with system size as $N^7$, $N^8$, $N^9$, and $N^{10}$, respectively, limiting their application primarily to small molecules.

Multi-reference methods such as complete active space self-consistent field (CASSCF), CASSCF with a second-order perturbation energy correction (CASPT2),~\cite{CASPT2} and the second-order $n$-electron valence state perturbation theory (NEVPT2)~\cite{NEVPT2} are better suited to treat double excitations. Unfortunately, these methods scale exponentially with the number of orbitals and electrons in the active space, limiting them to small active spaces and system size. CASPT2 tends to underestimate the VTEs of organic molecules and a shift is generally introduced in the zeroth-order Hamiltonian to provide better global agreement.~\cite{IPEA} These multi-reference methods also rely on chemical intuition to choose which orbitals to include in the active space, but it is often unintuitive which active space will capture the important determinants of a given state.

Selected configuration interaction (sCI)~\cite{CIPSI1,CIPSI2} methods are capable of obtaining double excitation energies and have recently been shown to reach full-CI (FCI) quality energies for small molecules.~\cite{CIPSI3} Among these approaches is the CI perturbatively selected iteratively (CIPSI)~\cite{CIPSI1} method in which determinants are selected based on their contribution to the second-order perturbation (PT2) energy, so that the most energetically relevant determinants are included in the determinant expansion first. This selection criterion both circumvents the excitation-ordered exponential expansion of the wave function used by CI and CC methods and removes the dependence on one's chemical intuition to decide which determinants to include in the expansion.

Quantum Monte Carlo (QMC) methods, specifically, variational (VMC) and fixed-node diffusion Monte Carlo (DMC) are promising first-principles approaches for solving the Schr\"{o}dinger equation stochastically. These methods scale favorably with system size ($N^4$ with $N$ the number of electrons) and naturally parallelize.~\cite{QMC1,QMC2,QMC3} In addition, recent improvements in QMC algorithms~\cite{QMC4,QMC6,QMC7,ammar_2022} allow for fast optimization of trial wave functions with thousands of parameters. When using determinant expansions provided by the CIPSI method fully optimized in the presence of a Jastrow factor as trial wave functions, VMC and DMC have been shown to provide accurate excitation energies.~\cite{QMC9,QMC10,dash2021} The role of the Jastrow factor is to account for dynamic correlations allowing for even shorter determinant expansions.~\cite{QMC11,QMC12,QMC13,QMC14,QMC15,QMC16,QMC17,QMC18}

It is shown here that the same VMC and DMC protocols used previously to calculate single excitations,~\cite{dash2021} can be used to treat double excitations just as precisely. The pure double excitations of nitroxyl, glyoxal, and tetrazine are calculated using VMC and DMC. In addition, a prediction is made for two excitation energies in cyclopentadienone which have strong double-excitation character in both excited states. The trial wave functions of the ground and excited states are optimized simultaneously, maintaining orthogonality on-the-fly by imposing an overlap penalty.




\section{\label{sec:methods}Methods}

In the following, we present how we build the trial wave functions for the QMC calculations, and how we optimize them in VMC energy minimization using a penalty-based, state-specific scheme for excited states, similar to previous approaches.~\cite{vmcpen1,vmcpen2,Wagner2021}

\subsection{Wave Functions}

The QMC trial wave functions take the Jastrow-Slater form,
\begin{equation}\label{WF}
\Psi_{I}=J_I\sum_{i=1}^{N_{\mathrm{CSF}}}c_{Ii}C_i[\left\{ \phi \right\}_{I}],
\end{equation}
where $C_i$ are spin-adapted configuration state functions (CSF), $c_{Ii}$ the expansion coefficients, $J_I$ the Jastrow factors, and $\left\{\phi\right\}_I$ the one-particle orbitals. In all calculations presented here, the ground and excited states have the same symmetry, so the CSFs, $C_i$, share the same orbital occupation patterns for all states. Note that each state, $I$, in Eq.~\ref{WF} has its own optimal Jastrow factor, set of orbitals, and expansion coefficients. When each state has its own optimizable Jastrow and orbital parameters, the approach is referred to as a state-specific optimization, as opposed to a state-average optimization where all states share a common Jastrow factor and set of orbitals and only the expansion coefficients are state-dependent.~\cite{QMC10,QMC11} 

The expansion coefficients and orbitals are initialized for the VMC optimization by one of two methods: (i) a $N_{\mathrm{state}}$-state-average [SA($N_{\mathrm{state}}$)] CASSCF calculation or (ii) a CIPSI calculation, which builds off of approach (i). In approach (i), the SA-CASSCF wave functions are used as starting determinantal components in the VMC optimization (results reported in SI). 
In approach (ii), one only utilizes the orbitals produced by the SA-CASSCF calculation as the molecular orbital basis for a CIPSI expansion; in the VMC optimization, the starting determinantal components include the CASSCF orbitals and CIPSI coefficients, $c_{Ii}$. 

Alternatively, in approach (ii), the natural orbitals (NO) are calculated from a first CIPSI expansion followed by a second CIPSI calculation in the NO basis. Performing a CIPSI expansion in the NO basis tends to lead to a smoother convergence to lower energies with fewer determinants~\cite{questdb2} compared to using a basis of SA-CASSCF orbitals. In other words, one can obtain a higher quality wave function with fewer determinants via relaxation of the molecular orbital basis. 

Further details on the CASSCF initial wave functions and CIPSI calculations using the NO basis are provided in the supplementary information (SI). 




\subsection{Penalty-Based State-Specific Method}\label{vmcmethod}

When trying to optimize a state which is not the lowest in its symmetry class, the state can collapse to a lower-energy eigenstate. One way to prevent this is to impose orthogonality between the higher energy state and all states lower in energy. To this end we employ a penalty-based state-specific method which requires minimizing the objective function,~\cite{Wagner2021}

\begin{equation}\label{eq:objfunc}
O_I[\Psi_I]=E_I[\Psi_I] + \sum_{J\ne I}\lambda_{IJ}|S_{IJ}|^2,
\end{equation}
where $E_I$ is given by,
\begin{equation}\label{eq:energy}
E_I=\frac{\langle \Psi_I | H | \Psi_I \rangle}{\langle \Psi_I | \Psi_I \rangle}=\langle  E_{\mathrm{L}}^I \left(\mathbf{R}\right) \rangle_{\mathbf{R}\sim |\Psi_I \left(\mathbf{R}\right)|^2}.
\end{equation}
$E_{\mathrm{L}}^I$ is the local energy, $H\Psi_I/\Psi_I$, and  $\langle \cdot \rangle_{\mathbf{R}\sim \cdot}$ denotes the Monte Carlo average of the quantity in brackets over the electron configurations, $\mathbf{R}$, sampled from the indicated distribution (in this case, $|\Psi_I \left(\mathbf{R}\right)|^2$). 
The normalized overlap, $S_{IJ}$, is given by,
\begin{equation}\label{eq:overlap}
S_{IJ}=\frac{\langle \Psi_I | \Psi_J \rangle}{\sqrt{\langle \Psi_I | \Psi_I \rangle \langle \Psi_J | \Psi_J \rangle}}.
\end{equation}

In order to simultaneously sample quantities for multiple states with comparable efficiency, a guiding function, $\Psi_{\mathrm{g}}=\sqrt{\sum_{I}|\Psi_I|^2}$, is introduced and the energy of a given state (Eq.~\ref{eq:energy}) is estimated as a weighted average
\begin{equation}\label{eq:sampled-energy}
E_I=\frac{\left\langle t_I E_{\mathrm{L}}^I \right\rangle_{\mathbf{R}\sim\rho_{\mathrm{g}}(\mathbf{R})}}{\left\langle t_I \right\rangle_{\mathbf{R}\sim\rho_{\mathrm{g}}(\mathbf{R})}},
\end{equation}
where now the configurations are sampled from the distribution, $\rho_{\mathrm{g}}=|\Psi_{\mathrm{g}}|^2$ and the weight $t_I$ is given
by $t_I=|\Psi_I/\Psi_g|^2$. Similar weighted averages are calculated for the
derivatives of each state (see SI).

The Jastrow, orbital, and CSF parameters of each state are optimized using the stochastic reconfiguration (SR)~\cite{SR1,SR2} method. The gradients used in the SR equations are found by taking the parameter derivatives of Eq.~\ref{eq:objfunc}, also derived in the SI.

In the implementation proposed by Pathak {\it et al.}, the state being optimized, $\Psi_I$, is forced to be orthogonal to fixed, pre-optimized, lower-energy states, or anchor states, $\Psi_J$. Each higher-energy state is then optimized consecutively, imposing orthogonality with all anchor states lower in energy. Each $\lambda_{IJ}$ in Eq.~\ref{eq:objfunc} controls the overlap penalty for the state currently being optimized, $\Psi_I$, due to each anchor state, $\Psi_J$, which is suggested to be on the order of, and larger than, the energy spacing between the states.

The method is implemented here such that all states are optimized at the same time, with orthogonality imposed on-the-fly. Since all states are changing from one SR step to the next, the overlap penalty must be imposed between all states. Therefore, `$\ne$' is used in the summation in Eq.~\ref{eq:objfunc} instead of `$<$' so that each state is kept orthogonal to all states, even to those higher in energy. Importantly, in circumstances where states are close in energy, their order may change throughout the optimization and, with `$\ne$' in Eq.~\ref{eq:objfunc},  the ordering of the states in energy does not need to be known in advance.
Finally, the use of correlated sampling in the concomitant optmization of all states reduces the length of the Monte Carlo simulations needed to acquire a given accuracy on the energy differences among the states.

\subsection{\label{sec:results:lambda}Analysis of $\lambda_{IJ}$'s effect on energies and VTEs}

We illustrate the effect of $\lambda_{IJ}$ on the optimization of the
3-state cyclopentadienone system with a small CIPSI expansion on CAS orbitals, performing
a total of 800 SR iterations
for different choices $\lambda^{(i)}$, each corresponding to
a different triplet of values ($\lambda_{12}$,$\lambda_{13}$,$\lambda_{23}$).
In general, we expect that if $\lambda_{IJ}$ are too small
some of the states may loose orthogonality and collapse onto each other; 
on the other hand, if they are too large, the fluctuations of 
the overlaps $S_{IJ}$ will be amplified, downgrading
the efficiency of the SR minimization because more sampling will be
required to pin down the optimal variational parameters at the desired 
level of statistical precision.

\begin{figure}[!htb]
\begin{center}
\includegraphics[width=\columnwidth]{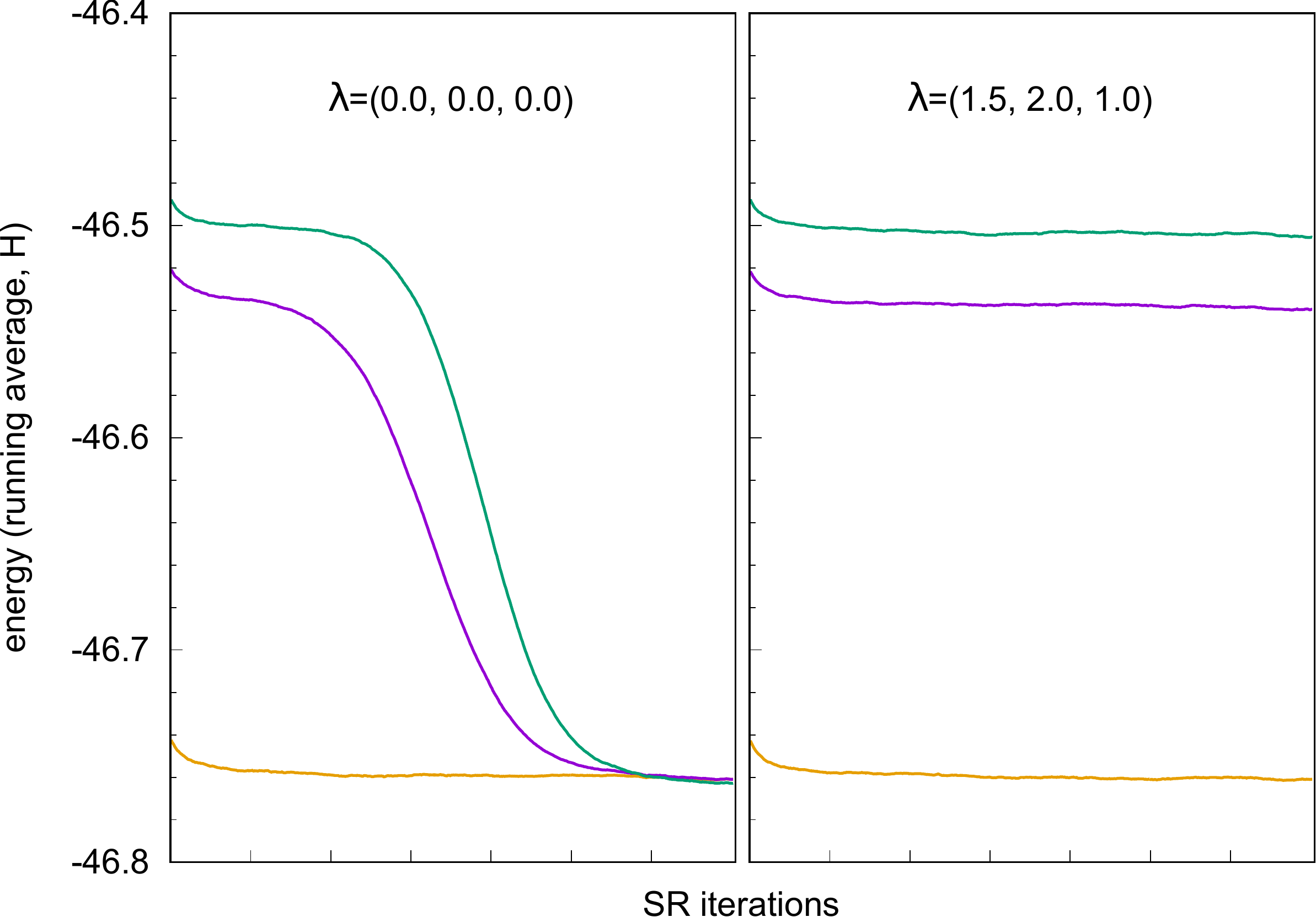}
\caption{\label{fig:cyclopentadienone-energy} Plot of the VMC energy during optimization of the 3-state cyclopentadienone system with a CIPSI wave function with 3036 determinants expanded on CAS orbitals. (Left) The collapse of the excited states to the ground state when $\lambda_{IJ}=0$. (Right) A stable optimization using $\lambda_{12}=1.5$, $\lambda_{13}=2.0$, and $\lambda_{23}=1.0$. The energy points in the plot are calculated from a running average of 100 SR iterations.}
\end{center}
\end{figure}

As shown in Figure~\ref{fig:cyclopentadienone-energy},
if all $\lambda_{IJ}$ are set to zero, the excited states 
collapse toward the ground-state wave function (left panel), 
whereas they are bounded by the respective eigenstates with the
choice of $\lambda_{IJ}$ indicated in the right panel.

\begin{figure}[!htb]
\begin{center}
\includegraphics[width=\columnwidth]{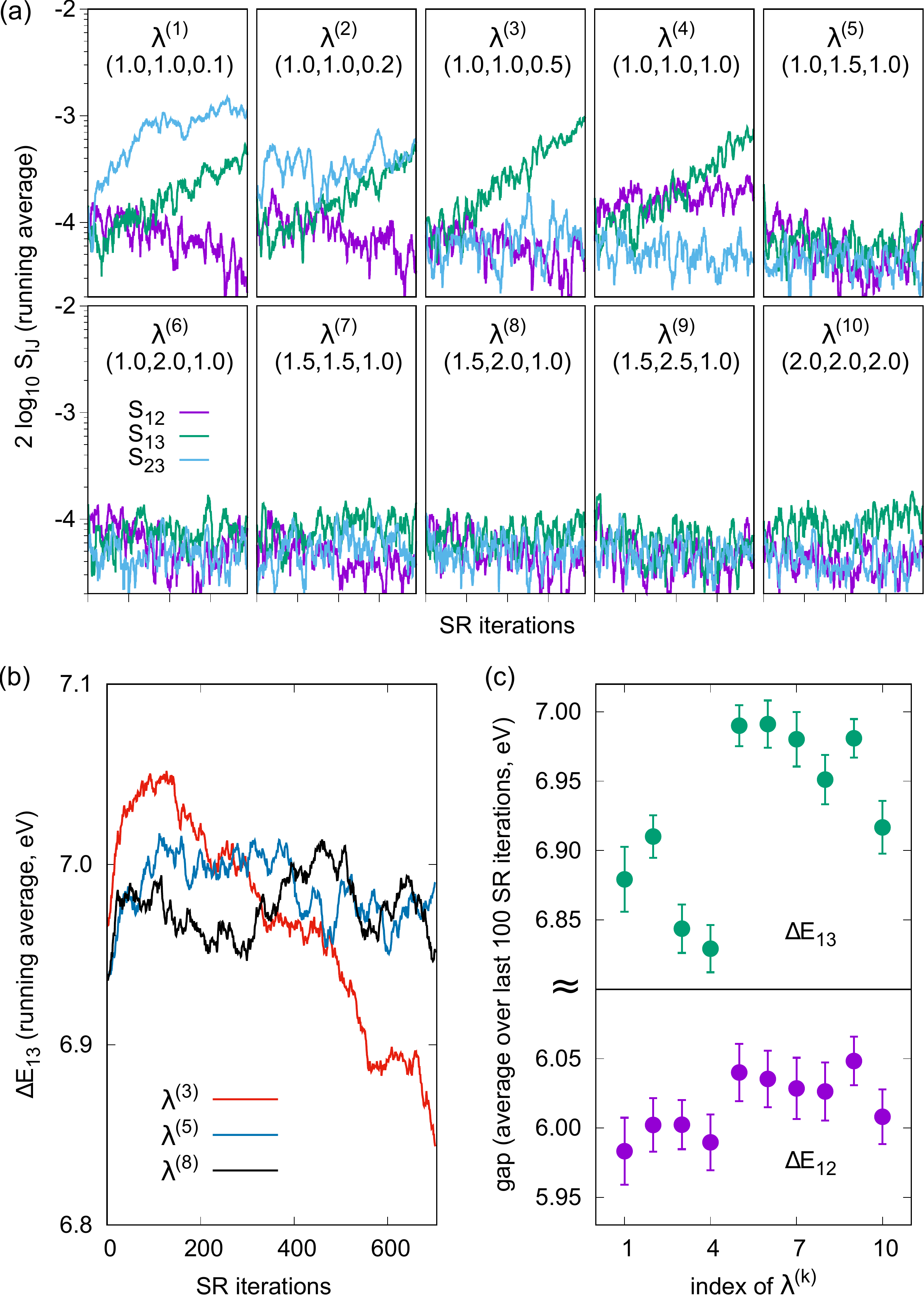}
\caption{\label{fig:cyclopentadienone-overlaps} Effect of $\lambda_{IJ}$ on the VMC optimization of the 3-state cyclopentadienone system with a CIPSI wave function with 3036 determinants expressed on CAS orbitals. (a) Logarithmic plot of $S^2_{IJ}$ at each SR iteration for different $\lambda^{(k)}=(\lambda_{12},\lambda_{13},\lambda_{23})$. (b) Running average (100 iterations) of $\Delta E_{13}$ for three different $\lambda^{(k)}$. (c) The excitations energies averaged over the last 100 SR iterations for each $\lambda^{(k)}$. }
\end{center}
\end{figure}

The selection of suitable values of $\lambda_{IJ}$ is neither difficult
nor critical. Figure~\ref{fig:cyclopentadienone-overlaps} shows 
the overlaps $|S_{IJ}|^2$ and excitation energies ($\Delta E_{IJ}$) for 
various choices of $\lambda_{IJ}$. In panel (a),
for $\lambda^{(1)-(4)}$, there is a noticeable increase in the overlaps, 
while the choices of $\lambda^{(5)-(9)}$ prevent the overlaps from increasing; 
however, for $\lambda^{(10)}$, $|S_{13}|^2$ tends to be uniformly larger. 
Panel (b) shows some examples of the correspondence between 
increasing (stable) overlap and decreasing (stable) gap.
Finally, panel (c) shows that all choices $\lambda^{(5)}$--$\lambda^{(9)}$
are equally good in terms of the final estimate of the excitation energies,
whereas for $\lambda^{(1)}$--$\lambda^{(4)}$ the gaps (particularly
$\Delta E_{13}$) are somewhat smaller and may eventually collapse 
with more SR iterations. In the case of $\lambda^{(10)}$, the signal
is stable along the iterations, but the excitation energy $E_{13}$ is
somewhat less accurate, than with $\lambda^{(5)}$--$\lambda^{(9)}$; 
it could be improved with more sampling per SR step, loosing however
efficiency.

In conclusion, there is a wide range of $\lambda_{IJ}$
($\lambda^{(5)}$--$\lambda^{(9)}$) where the estimate of the 
excitation energies is stable, consistent and efficient.


In general, the nature of each individual excited state
as well as the trial function itself will impact its sensitivity to the
value of $\lambda_{IJ}$. For instance, 
in the nitroxyl test case with the 2-determinant wave function (see SI), 
even with $\lambda_{IJ}=0$, the excited state
still cannot collapse completely to the ground state due to
orbital symmetries. 




\section{\label{sec:computational_details}Computational Details}

The geometries of nitroxyl, glyoxal, tetrazine, and cyclopentadienone are obtained from  ground-state optimizations at the level of CC3/aug-cc-pVTZ.~\cite{loos2019,questdb} In all calculations we utilize scalar-relativistic energy-consistent Hartree Fock (HF) pseudopotentials~\cite{BFD07} with the corresponding aug-cc-pVDZ Gaussian basis sets. The exponents of the diffuse functions are taken from the corresponding all-electron
Dunning's correlation-consistent basis sets~\cite{kendall1992}. For glyoxal, we also tested the use of an aug-cc-pVTZ basis set and found compatible excitation energies both at the VMC and DMC level. All HF and SA-CASSCF calculations are performed in GAMESS(US)~\cite{GAMESS} with equal weights on all states.

CIPSI calculations are performed with the Quantum Package~\cite{QP,QP2} program. All states of interest are singlets so determinants are chosen such that the expansions remain eigenstates of $\hat{S}^2$ with eigenvalue 0. For each molecule, the ground and excited states of interest have the same symmetry so the same set of orbitals are used in all state's determinant expansions. States are weighted and determinants are added to the CIPSI expansion such that the PT2 energy and variance of all states remain similar as this has been shown to give more accurate QMC excitation energies.~\cite{QMC9,QMC10,dash2021} The choice of weights in the CIPSI selection criterion is detailed for each molecule in the SI.

QMC calculations are performed with the CHAMP~\cite{CHAMP} program using the method detailed in Section~\ref{vmcmethod}. Damping parameters ranging from $\tau_{\mathrm{SR}}=$0.05$-$0.025~a.u. are used in the SR method during the VMC optimization. A low-memory conjugate-gradient algorithm~\cite{QMC5} is used to solve for the wave function parameters in the SR equations. All DMC calculations use a time-step of $\tau_{\mathrm{DMC}}=$0.02~a.u.

A value of $\lambda_{IJ}=1.0$~a.u. is used for all 2-state calculations while multiple values are used for the 3-state system as discussed in the previous Section and in the SI.

\section{\label{sec:results}Results}

\subsection{\label{sec:results:nitroxyl}Nitroxyl $1 ^1\mathrm{A'} \rightarrow 2 ^1\mathrm{A'}$ Double Excitation}

\begin{table*}[!htb]
\caption{Nitroxyl VMC and DMC total energies (a.u.) and excitation energies (eV). The number of determinants (\#~det) and parameters (\#~parm) in each trial wave function are also listed. The CIPSI expansions are performed on a NO basis and fully reoptimized in VMC.} 
\begin{tabular}{lrrllllll}
\hline
 & & & \multicolumn{3}{c}{VMC} & \multicolumn{3}{c}{DMC} \\
\cline{4-6}
\cline{7-9}
\multicolumn{1}{c}{WF} & \multicolumn{1}{c}{\# det} & \multicolumn{1}{c}{\# parm} & \multicolumn{1}{c}{$E\left(1 ^1\mathrm{A'}\right)$} & \multicolumn{1}{c}{$E\left(2 ^1\mathrm{A'}\right)$} & \multicolumn{1}{c}{$\Delta E$} & \multicolumn{1}{c}{$E\left(1 ^1\mathrm{A'}\right)$} & \multicolumn{1}{c}{$E\left(2 ^1\mathrm{A'}\right)$} & \multicolumn{1}{c}{$\Delta E$} \\
\hline
CIPSI &    321 &  850 & -26.4917(2) & -26.3322(2) & 4.34(1) & -26.5164(2) &
           -26.3583(2) & 4.30(1) \\
           &   1573 & 1446 & -26.5001(2) & -26.3415(2) & 4.31(1) & -26.5220(2) & -26.3636(2) & 4.31(1) \\
           &   2900 & 2002 & -26.5040(2) & -26.3447(2) & 4.34(1) & -26.5240(2) & -26.3652(2) & 4.32(1) \\
           &   7171 & 4124 & -26.5102(2) & -26.3498(2) & 4.36(1) & -26.5267(2) & -26.3677(2) & 4.33(1) \\
           &  10690 & 5712 & -26.5113(2) & -26.3519(2) & 4.34(1) & -26.5273(2) & -26.3686(2) & 4.32(1) \\
\hline
\multicolumn{8}{l}{TBE$^a$}       & \multicolumn{1}{l}{4.33} \\
\multicolumn{8}{l}{CC3/AVQZ$^b$} & \multicolumn{1}{l}{5.23} \\
\multicolumn{8}{l}{CC4/AVDZ$^c$}               & \multicolumn{1}{l}{4.454} \\
\multicolumn{8}{l}{CC4/AVTZ$^c$}               & \multicolumn{1}{l}{4.380} \\
\multicolumn{8}{l}{CCSDT/AVDZ$^b$}         & \multicolumn{1}{l}{4.756} \\
\multicolumn{8}{l}{CCSDTQ/AVDZ$^b$}        & \multicolumn{1}{l}{4.424} \\
\multicolumn{8}{l}{CCSDTQ/AVTZ$^c$}            & \multicolumn{1}{l}{4.364} \\
\multicolumn{8}{l}{exFCI/AVQZ$^d$}         & \multicolumn{1}{l}{4.32(0)} \\
\multicolumn{8}{l}{CASPT2(12,9)/AVQZ$^d$}  & \multicolumn{1}{l}{4.34} \\
\multicolumn{8}{l}{PC-NEVPT2(12,9)/AVQZ$^d$} & \multicolumn{1}{l}{4.35} \\
\hline
\multicolumn{9}{l}{$^a$Value is `safe', as defined in Ref.~\citenum{questdb}. FCI/AVTZ}\\
\multicolumn{9}{l}{$^b$Ref.~\citenum{loos2021}.}\\
\multicolumn{9}{l}{$^c$Ref.~\citenum{newcc4}.}\\
\multicolumn{9}{l}{$^d$Ref.~\citenum{loos2019}.}\\
\end{tabular}
\label{tab:nitroxyl}
\end{table*}

This transition in nitroxyl is a pure double $(n,n)\rightarrow(\pi^*,\pi^*)$ excitation. The VMC and DMC energies and VTEs are shown in Table~\ref{tab:nitroxyl}. There are reliable benchmark calculations for the $1 ^1\mathrm{A'} \rightarrow 2^1\mathrm{A'}$ VTE in nitroxyl. 
In particular, the exFCI/aug-cc-pVQZ (AVQZ) calculation predicts the lowest value for this excitation at 4.32~eV. 
The latest~\cite{newcc4} CC4 and CCSDTQ calculations using the aug-cc-pVTZ (AVTZ) basis set somewhat overestimate it at 4.380 and 4.364~eV, respectively.


\begin{figure}[htb]
\begin{center}
\includegraphics[width=\columnwidth]{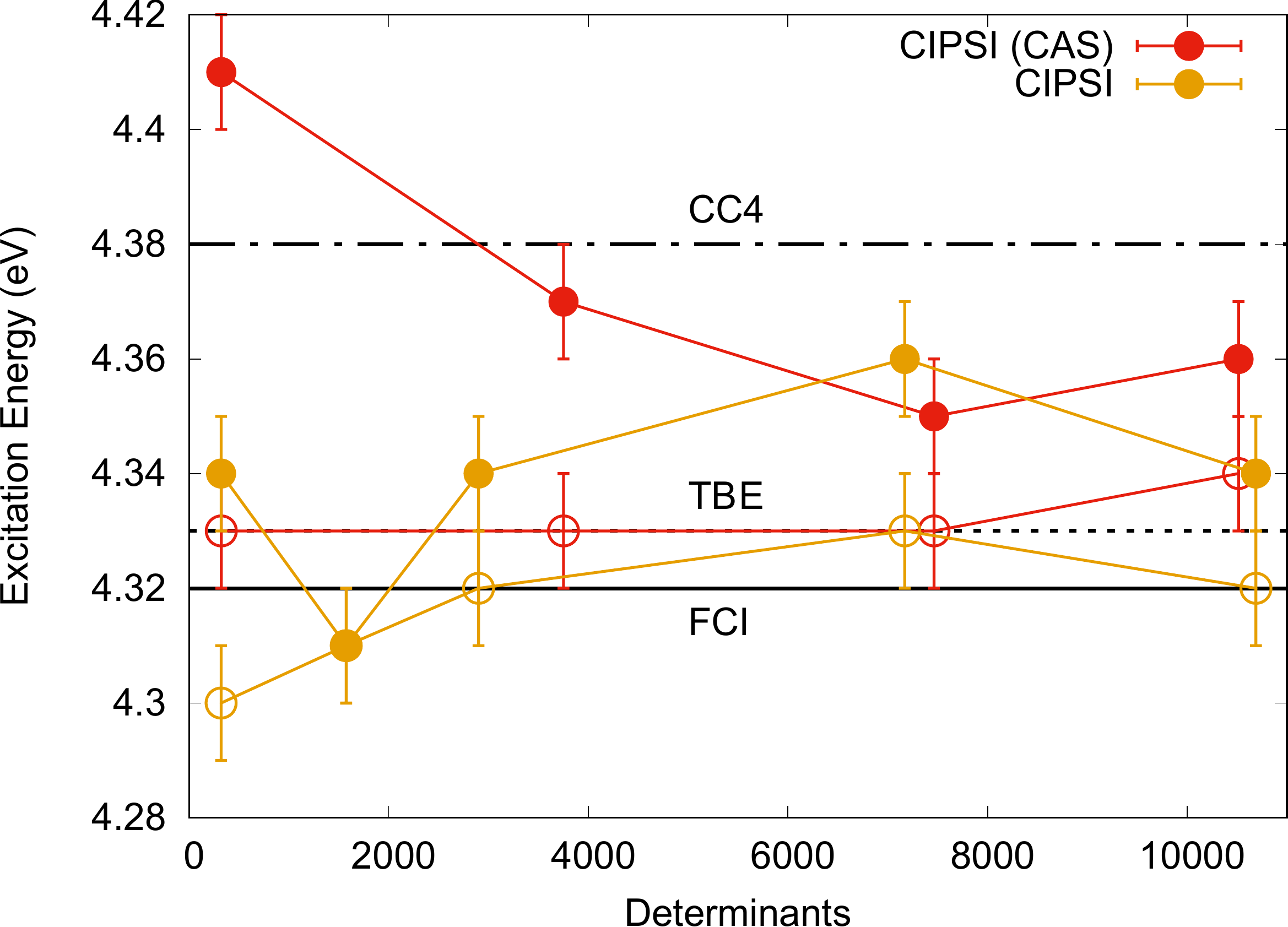}
\caption{\label{fig:nitroxyl-cipsi-all-vmc-dmc} Nitroxyl VMC (filled points) and DMC (empty points) excitation energies for different CIPSI wave functions expanded either on starting CAS [SA(2)-CASSCF(12,9)] or NO basis, and fully reoptimized in VMC. Horizontal lines represent reference values. TBE: FCI/ATVZ, FCI: exFCI/AVQZ, CC4: CC4/AVTZ. }
\end{center}
\end{figure}


CIPSI wave functions are expanded in a CAS orbital basis [SA(2)-CASSCF(12,9)] as well as in an NO basis, while matching the PT2 energies and variance. Upon state-specific optimization of the Jastrow-CIPSI wave functions, the VMC and DMC VTEs are very similar for different matching and orbital basis (see Figure~~\ref{fig:nitroxyl-cipsi-all-vmc-dmc} and SI) and all are in very good agreement with benchmarks. In addition, consistency of VMC with the DMC excitation energy is reached at fairly small determinant expansions ($\sim$2,000). Due to the fast convergence of the VMC and DMC excitation energies, the use of NOs in the CIPSI expansion is not necessary, but still provides similar results.

Previous results for nitroxyl show that at least CC4 and CCSDTQ is necessary to capture the important correlations of the $1 ^1\mathrm{A'} \rightarrow 2^1\mathrm{A'}$ double excitation. There is also a strong basis set dependence on the excitation with CC methods, yet the VMC and DMC calculations starting from AVDZ with pseudopotentials, reach agreement with expensive quantum chemistry calculations which require the use of at least AVTZ basis set.

\subsection{\label{sec:results:glyoxal}Glyoxal $1 ^1\mathrm{A}_{\mathrm{g}} \rightarrow 2 ^1\mathrm{A}_{\mathrm{g}}$ Double Excitation}

\begin{table*}[!htb]
\caption{Glyoxal VMC and DMC total energies (a.u.) and excitation energies (eV). The number of determinants (\#~det) and parameters ( \#~parm) in each trial wave function are also presented. The CIPSI expansions are performed on a NO basis and fully reoptimized in VMC.}
\begin{tabular}{lrrllllll}
\hline
 & & & \multicolumn{3}{c}{VMC} & \multicolumn{3}{c}{DMC} \\ 
\cline{4-6}
\cline{7-9}
\multicolumn{1}{c}{WF} & \multicolumn{1}{c}{\# det} & \multicolumn{1}{c}{\# parm} & \multicolumn{1}{c}{$E\left(1 ^1\mathrm{A_g}\right)$} & \multicolumn{1}{c}{$E\left(2 ^1\mathrm{A_g}\right)$} & \multicolumn{1}{c}{$\Delta E$} & \multicolumn{1}{c}{$E\left(1 ^1\mathrm{A_g}\right)$} & \multicolumn{1}{c}{$E\left(2 ^1\mathrm{A_g}\right)$} & \multicolumn{1}{c}{$\Delta E$} \\
\hline
CIPSI                   &  3749 & 2908 & -44.6072(2) & -44.3957(2) & 5.76(1)   & -44.6564(2)   & -44.4482(2)   & 5.67(1) \\
                        &  5046 & 3498 & -44.6127(2) & -44.4004(2) & 5.78(1)   & -44.6600(2)   & -44.4509(2)   & 5.69(1) \\
                        &  7040 & 4500 & -44.6150(2) & -44.4048(2) & 5.72(1)   & -44.6613(2)   & -44.4534(2)   & 5.66(1) \\
                        &  8642 & 5433 & -44.6181(2) & -44.4098(2) & 5.67(1)   & -44.6632(2)   & -44.4559(2)   & 5.64(1) \\
                        & 10010 & 6188 & -44.6205(2) & -44.4114(2) & 5.69(1)   & -44.6635(2)   & -44.4567(2)   & 5.63(1) \\
                        & 12132 & 6269 & -44.6237(2) & -44.4145(2) & 5.70(1)   & -44.6654(2)   & -44.4576(2)   & 5.66(1) \\
                        & 15349 & 8577 & -44.6276(2) & -44.4180(2) & 5.70(1)   & -44.6670(2)   & -44.4600(2)   & 5.63(1) \\
\hline
\multicolumn{8}{l}{TBE$^a$}         &  \multicolumn{1}{l}{5.61} \\
\multicolumn{8}{l}{CC3/AVQZ$^b$}            &  \multicolumn{1}{l}{6.76} \\
\multicolumn{8}{l}{CC4/6-31+G*$^c$}           &  \multicolumn{1}{l}{5.699} \\
\multicolumn{8}{l}{CC4/AVDZ$^c$}              &  \multicolumn{1}{l}{5.593} \\
\multicolumn{8}{l}{CCSDTQ/6-31+G*$^c$}        &  \multicolumn{1}{l}{5.670} \\
\multicolumn{8}{l}{exFCI/AVDZ}    &  \multicolumn{1}{l}{5.56(11)} \\
\multicolumn{8}{l}{SAC-CI/AVDZ$^d$}    &  \multicolumn{1}{l}{5.66} \\
\multicolumn{8}{l}{CASPT2(14,12)/AVQZ$^b$}    &  \multicolumn{1}{l}{5.43} \\
\multicolumn{8}{l}{PC-NEVPT2(14,12)/AVQZ$^b$} &  \multicolumn{1}{l}{5.52} \\
\hline
\multicolumn{9}{l}{$^a$This value is `safe', as defined in Ref.~\citenum{questdb}. exFCI/AVDZ + (CCSDT/AVTZ-CCSDT/AVDZ) } \\
\multicolumn{9}{l}{$^b$Ref.~\citenum{loos2019}.} \\
\multicolumn{9}{l}{$^c$Ref.~\citenum{newcc4}.} \\
\multicolumn{9}{l}{$^d$Ref.~\citenum{SACCI}. See reference for basis set details.} \\
\end{tabular}
\label{tab:glyoxal}
\end{table*}

The pure double excitation in glyoxal also corresponds to a $(n,n)\rightarrow(\pi^*,\pi^*)$ transition. Since the system size is larger compared to nitroxyl, the same level of CC4 and CCSDTQ calculations is not available, and there appears to be a large basis set dependence on the most recent CC4 data.~\cite{newcc4} A symmetry adapted cluster CI (SAC-CI) calculation of this excitation is available~\cite{SACCI} (5.66~eV) and agrees well with the recent TBE, which is a basis set corrected FCI/AVDZ value of 5.61~eV.~\cite{questdb} These values also agree well with the highest level CCSDTQ estimate for this VTE, 5.670~eV,~\cite{newcc4} all shown and compared to our QMC results in Table~\ref{tab:glyoxal}.

\begin{figure}[htb]
\begin{center}
\includegraphics[width=\columnwidth]{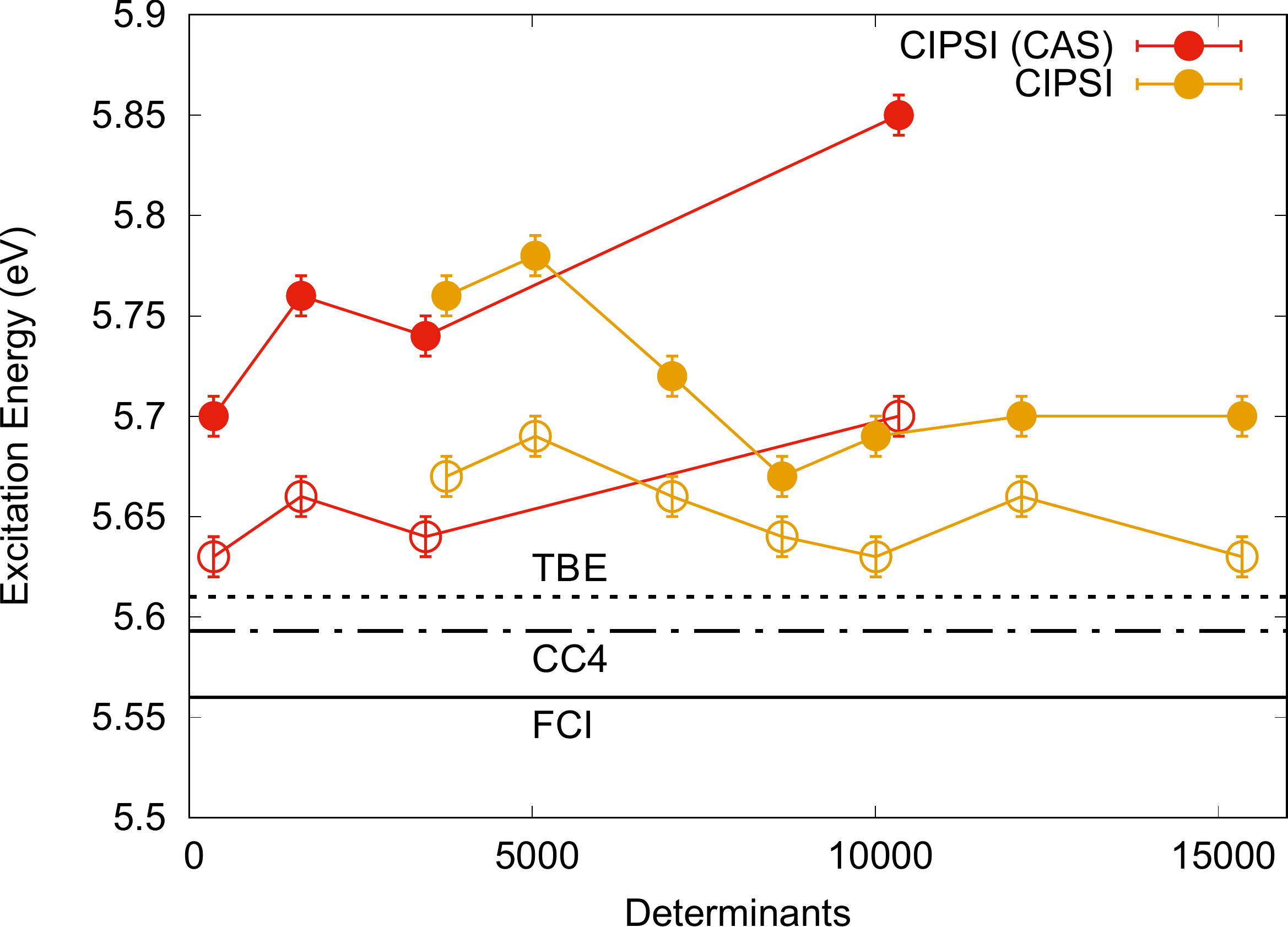}
\caption{\label{fig:glyoxal-cipsi-all-vmc-dmc} Glyoxal VMC (filled circles) and DMC (empty circles) excitations energies for different CIPSI wave functions expanded either on starting CAS [SA(2)-CASSCF(14,12)] or NO basis and fully reoptimized in VMC. Horizontal lines represent reference values. TBE: FCI/AVDZ + (CCSDT/AVTZ - CCSDT/AVDZ), CC4: CC4/AVDZ, FCI: exFCI/AVDZ (error bar of $\pm$0.11~eV). }
\end{center}
\end{figure}


PT2 and variance-matched CIPSI expansions are generated from the optimal orbitals of a SA(2)-CASSCF(14,12) calculation.
As shown in Figure~\ref{fig:glyoxal-cipsi-all-vmc-dmc}, after the optimization of the CIPSI (CAS) wave functions, we observe a somewhat difficult convergence of the excitation energy as the determinant expansion is increased from $\sim400$ to $\sim10,000$. The DMC values tend to also increase, although less dramatically.

When using NOs built from modest-sized CIPSI expansions ($10^4$ determinants) and re-performing the CIPSI expansions in the presence of the Jastrow factor, the VMC and DMC excitation energies appear to converge nicely and to a value in agreement with available benchmarks. It is possible that still larger CIPSI expansions and more rounds of NO generation could bring the VMC into better alignment with the DMC, which differs at most by $0.05$~eV. The DMC value for the largest determinant expansion in the NO basis is 5.63(1)~eV, which is $\sim$0.01~eV from the current TBE value and $\sim$0.02~eV from the SAC-CI/AVDZ and CC4/AVDZ values. This close agreement and the consistency of the DMC results with different number of determinants in the CIPSI wave functions suggests the value of 5.63(1)~eV to be a reasonable prediction for this excitation.

\subsection{\label{sec:results:tetrazine}Tetrazine $1 ^1\mathrm{A}_{\mathrm{1g}} \rightarrow 2 ^1\mathrm{A}_{\mathrm{1g}}$ Double Excitation}

\begin{table*}[!htb]
\caption{\label{tetrazine-table}Tetrazine VMC and DMC total energies (a.u.) and excitation energies (eV). The number of determinants (\#~det) and parameters ( \#~parm) in each trial wave function are also presented. The CIPSI expansions are performed on a NO basis and fully optimized in VMC. }
\begin{tabular}{lrrllllll}
\hline
 & & & \multicolumn{3}{c}{VMC} & \multicolumn{3}{c}{DMC} \\ 
\cline{4-6}
\cline{7-9}
\multicolumn{1}{c}{WF} & \multicolumn{1}{c}{\# det} & \multicolumn{1}{c}{\# parm} & \multicolumn{1}{c}{$E\left(1 ^1\mathrm{A_{1g}}\right)$} & \multicolumn{1}{c}{$E\left(2 ^1\mathrm{A_{1g}}\right)$} & \multicolumn{1}{c}{$\Delta E$} & \multicolumn{1}{c}{$E\left(1 ^1\mathrm{A_{1g}}\right)$} & \multicolumn{1}{c}{$E\left(2 ^1\mathrm{A_{1g}}\right)$} & \multicolumn{1}{c}{$\Delta E$} \\
\hline
CIPSI           &  3572 & 2414 & -52.2198(2) & -52.0346(2) & 5.04(1)  & -52.2935(3)   & -52.1098(3)   & 5.00(1) \\
                     &  7025 & 4141 & -52.2300(2) & -52.0451(2) & 5.03(1)  & -52.2981(3)   & -52.1148(3)   & 4.99(1) \\
                     & 10723 & 6053 & -52.2379(2) & -52.0528(2) & 5.04(1)  & -52.3015(3)   & -52.1182(3)   & 4.99(1) \\
\hline
\multicolumn{8}{l}{TBE$^a$}                  &  \multicolumn{1}{l}{4.61} \\
\multicolumn{8}{l}{CC3/6-31+G*$^b$}                 &  \multicolumn{1}{l}{6.22} \\
\multicolumn{8}{l}{CC3/AVQZ$^b$}                 &  \multicolumn{1}{l}{6.19} \\
\multicolumn{8}{l}{CC4/6-31+G*$^c$}              &  \multicolumn{1}{l}{5.06} \\
\multicolumn{8}{l}{CCSDT/AVTZ$^b$}               &  \multicolumn{1}{l}{5.96} \\
\multicolumn{8}{l}{exFCI/AVDZ}               &  \multicolumn{1}{l}{5.15(3)} \\
\multicolumn{8}{l}{CASPT2(14,10)/AVQZ$^b$}       &  \multicolumn{1}{l}{4.68} \\
\multicolumn{8}{l}{PC-NEVPT2(14,10)/AVQZ$^b$}    &  \multicolumn{1}{l}{4.60} \\
\hline
\multicolumn{9}{l}{$^a$This value is `unsafe', as defined in Ref~\citenum{questdb}. NEVTPT2/AVTZ} \\
\multicolumn{9}{l}{$^b$Ref.~\citenum{loos2019}.} \\
\multicolumn{9}{l}{$^c$Ref.~\citenum{newcc4}.} \\
\end{tabular}
\label{tab:tetrazine}
\end{table*}

The tetrazine ($s$-tetrazine) pure double excitation of interest is another $(n,n)\rightarrow(\pi^*,\pi^*)$ transition. There are few reliable benchmark calculations available for this transition due to the molecule's size in addition to its genuine double nature. 
Although CC3 does not properly treat double excitations, calculations using 6-31+G* up to AVQZ show that there is very little basis set effect, only a change of $\sim0.03$~eV on the excitation energy. If the CC4/6-31+G* value of 5.06~eV is just as similar for larger basis sets, then this value can be seen as a good guess for the VTE. In any case, without the calculations available to be sure, the VMC and DMC presented in Table~\ref{tetrazine-table} suggest a value close to 5.0~eV for the VTE.

\begin{figure}[htb]
\begin{center}
\includegraphics[scale=0.35]{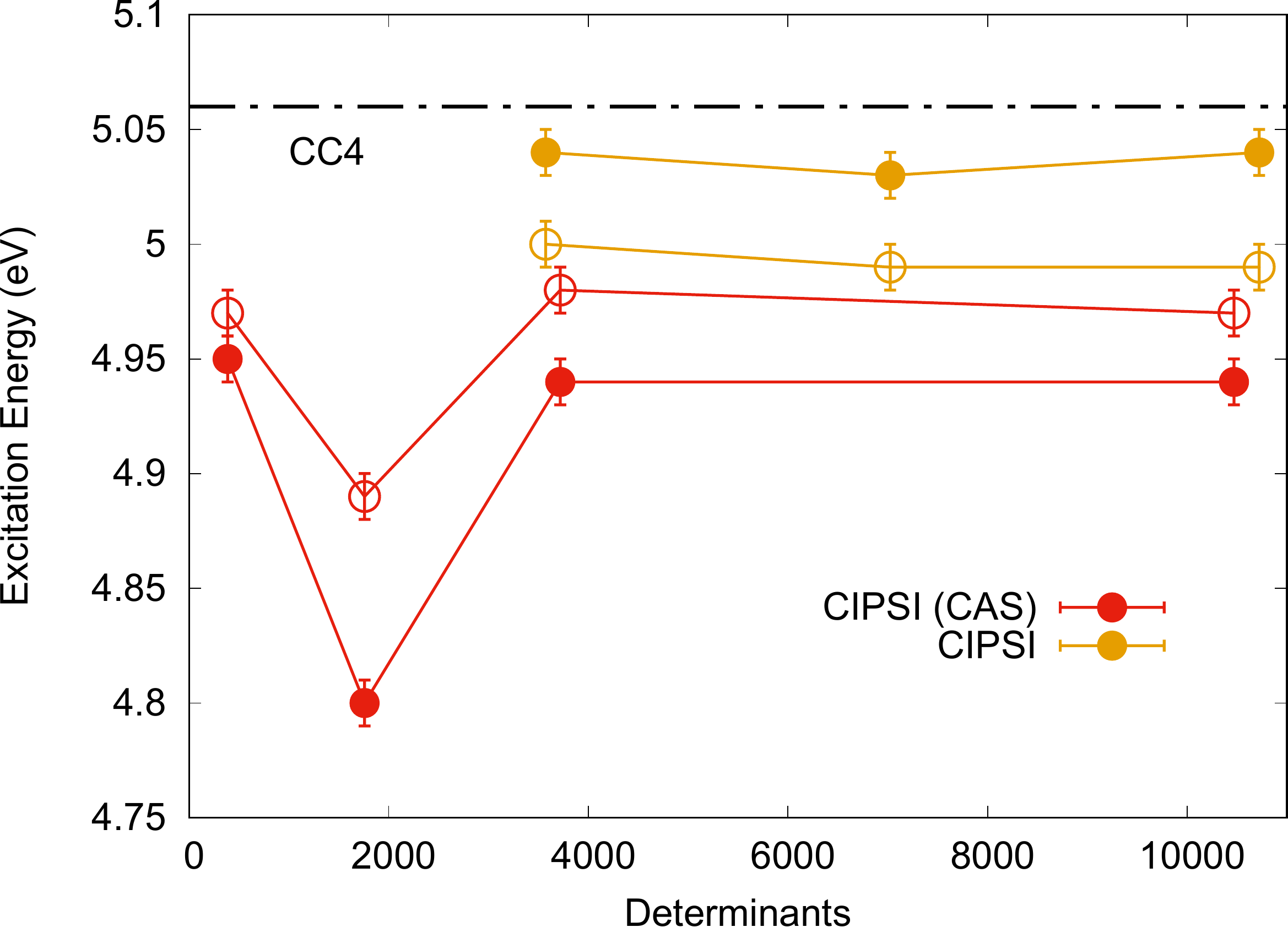}
\caption{\label{fig:tetrazine-cipsi-all-vmc-dmc} Tetrazine VMC (filled circles) and DMC (empty circles) excitations energies for different CIPSI wave functions expanded either on starting CAS [SA(2)-CASSCF(14,10)] or NO basis, and fully reoptimized in VMC. Horizontal lines represent the reference value. CC4: CC4/6-31+G*. }
\end{center}
\end{figure}


The VMC optimization of the CAS orbital-based [SA(2)-CASSCF(14,10)] CIPSI expansions tends towards 4.94(1)~eV (see SI) while the NO based CIPSI goes to 5.04(1)~eV, both converging after about 3,600$-$3,700 determinants. While the converged VMC excitations energies are not quite in agreement, their DMC values are consistent, with VTEs of 4.97(1) and 4.99(1)~eV at the largest determinant expansions (see Figure~\ref{fig:tetrazine-cipsi-all-vmc-dmc}). Even more so than in glyoxal, the consistency of the DMC value for the CIPSI expansions of different length and initial MO basis suggests it should be a trusted benchmark for this transition.


\subsection{\label{sec:results:cyclopentadienone}3-State VMC Optimization: Cyclopentadienone $1^1\mathrm{A}_1 \rightarrow 2^1\mathrm{A}_1$ and $1^1\mathrm{A}_1 \rightarrow 3^1\mathrm{A}_1$ Excitations}

\begin{table*}[!htb]
\caption{Cyclopentadienone VMC and DMC excitation energies (eV). The number of determinants (\#~det) and parameters (\#~parm) in each trial wave function is also listed. The CIPSI expansions are performed on a NO basis and fully optimized in VMC. Reference values for the double and single excitation are placed in the $\Delta E_{12}$ and $\Delta E_{13}$ columns, respectively, to match the order of the states calculated with VMC and DMC. The VMC optimization of the two smaller CIPSIs used $\lambda_{IJ}=$(1.5,2.0,1.0) and, of the two larger, (1.0,1.0,1.0).}
\begin{tabular}{lrrllllll}
\hline
 & & & \multicolumn{3}{c}{VMC} & \multicolumn{3}{c}{DMC} \\ 
\cline{4-6}
\cline{7-9}
\multicolumn{1}{c}{WF} & \multicolumn{1}{c}{\# det} & \multicolumn{1}{c}{\# parm} & \multicolumn{1}{c}{$\Delta E_{12}$} & \multicolumn{1}{c}{$\Delta E_{13}$} & \multicolumn{1}{c}{$\Delta E_{23}$} & \multicolumn{1}{c}{$\Delta E_{12}$} & \multicolumn{1}{c}{$\Delta E_{13}$} & \multicolumn{1}{c}{$\Delta E_{23}$} \\
\hline
CIPSI  &  1025 &  2403 & 5.94(1) & 6.92(1) & 0.98(1) & 5.90(1) & 6.87(1) & 0.97(1) \\
            &  3033 &  3645 & 5.95(1) & 6.96(1) & 1.01(1) & 5.87(1) & 6.87(1) & 1.00(1) \\
            &  7151 &  5365 & 6.03(1) & 7.05(1) & 1.02(1) & 5.92(1) & 6.91(1) & 0.99(1) \\
            & 10206 &  6574 & 6.04(1) & 7.03(1) & 0.99(1) & 5.90(1) & 6.89(1) & 0.99(1) \\
\hline
\multicolumn{6}{l}{TBE$^a$} & \multicolumn{1}{l}{6.00$^b$} & \multicolumn{1}{l}{6.09$^c$} & \multicolumn{1}{l}{0.09} \\
\multicolumn{6}{l}{ADC(3)/AVTZ}  & \multicolumn{1}{l}{4.59$^d$} & \multicolumn{1}{l}{6.50$^d$} & \multicolumn{1}{l}{1.91} \\
\multicolumn{6}{l}{CC2/AVTZ}     &              & \multicolumn{1}{l}{6.50$^d$} & \multicolumn{1}{l}{} \\
\multicolumn{6}{l}{CC3/AVTZ}     & \multicolumn{1}{l}{7.10$^d$} & \multicolumn{1}{l}{6.21$^d$} & \multicolumn{1}{l}{-0.89} \\
\multicolumn{6}{l}{CCSD/AVTZ}    &              & \multicolumn{1}{l}{6.68$^d$} & \multicolumn{1}{l}{} \\
\multicolumn{6}{l}{CCSDT-3/AVTZ} &              & \multicolumn{1}{l}{6.33$^d$} & \multicolumn{1}{l}{} \\
\multicolumn{6}{l}{exFCI/AVDZ}   & \multicolumn{1}{l}{5.81(39)} & \multicolumn{1}{l}{6.93(24)} & \multicolumn{1}{l}{1.10(13)} \\
\hline
\multicolumn{9}{l}{$^a$This value is `unsafe' as defined in Ref.~\citenum{questdb}.} \\
\multicolumn{9}{l}{$^b$NEVPT2/AVTZ.} \\
\multicolumn{9}{l}{$^c$CCSDT/AVDZ + ( CC3/AVTZ-CC3/AVDZ ).} \\
\multicolumn{9}{l}{$^d$Values are also `unsafe'.} \\
\end{tabular}
\label{tab:cyclopentadienone-dE}
\end{table*}

To show the capabilities of the VMC optimization method, the three lowest $^1\mathrm{A}_1$ states of cyclopentadienone are optimized simultaneously to determine the two lowest excitation energies. One is deemed a double excitation ($\pi$,$\pi\rightarrow \pi^*$,$\pi^*$) with $\sim$50\%~$T_1$, and the other a single excitation ($\pi\rightarrow \pi^*$) with $\sim$74\%~$T_1$ by CC3 calculations. All results are collected in Table~\ref{tab:cyclopentadienone-dE}.

\begin{figure}[htb]
\begin{center}
\includegraphics[width=\columnwidth]{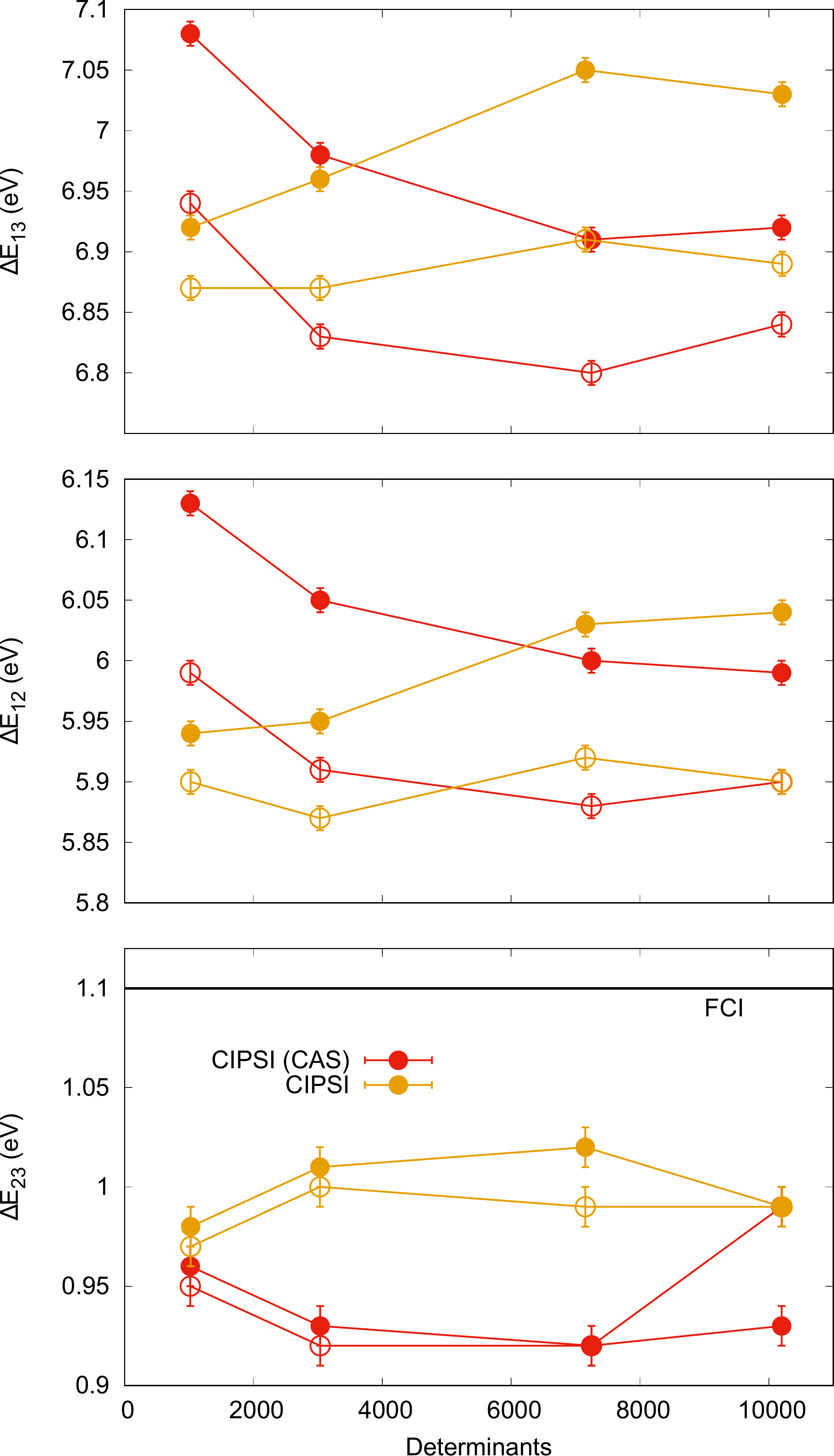}
\caption{\label{fig:cyclopentadienone-cipsi-all-vmc-dmc} Cyclopentadienone VMC (filled circles) and DMC (empty circles) excitations energies for different CIPSI wave functions expanded either on starting CAS [SA(3)-CASSCF(6,6)] or NO basis, and fully optimized in VMC. FCI: exFCI/AVDZ (error bar of $\pm$0.13~eV).  }
\end{center}
\end{figure}

CC3/AVTZ predicts a value of 7.10~eV for the double excitation and 6.21~eV for the single excitation, which is the opposite of what is found here. The CC3 value for the double excitation is not trustable since it misses quadruple excitations important to describing this type of excitation. Veril et al.~\cite{questdb} determine the CC3 and CCSDT calculations of these excitations to be `unsafe' since the they do not agree to within 0.03$-$0.04~eV. The relatively low $T_1$ of the single excitation could lead to a poor treatment by CC3 and CCSDT. No CC4 or CCSDTQ calculations are currently available for reliable comparison to the present results for the double excitation energy, and all benchmark VTEs in Table~\ref{tab:cyclopentadienone-dE} are considered unsafe. An exFCI/AVDZ calculation is performed in the present work, although it has fairly large error bars.


All VMC and DMC performed on CIPSI wave functions give roughly the same results (see Table~\ref{tab:cyclopentadienone-dE}, Figure~\ref{fig:cyclopentadienone-cipsi-all-vmc-dmc}, and SI). The largest CIPSI wave function in the NO basis has DMC $\Delta E_{12}$, $\Delta E_{13}$, and $\Delta E_{23}$ values of 5.90(1)~eV, 6.89(1)~eV, and 0.99~eV, respectively. 

The present calculations definitively show the ordering of the two lowest $\mathrm{^1A_1}$ excited states. The largest CIPSI calculation performed in the CAS orbital basis [SA(3)-CASSCF(6,6)] finds the $\mathrm{2 ^1A_1}$ state to have coefficients of 0.70 and 0.46 for the dominant doubly [$(\pi,\pi)\rightarrow (\pi^*,\pi^*)$] and singly excited ($\pi\rightarrow \pi^*$) CSFs, respectively, while the $\mathrm{3 ^1A_1}$ state has coefficients of 0.49 and 0.75. The coefficients show a clear mixing of singly and doubly excited determinants in both states (also suggested by the $T_1$ values), but the $\mathrm{2 ^1A_1}$ state is clearly distinguishable as the double excitation and the $\mathrm{3 ^1A_1}$ state as the single. 

\section{\label{sec:conclusions}Conclusions}

With the use of a penalty-based, state-specific optimization scheme, we show that the same QMC method which provides accurate single-excitations energies can also be applied to double-excitations. That is, we use PT2 energy and variance matched CIPSI determinant expansions as the Slater part of the Jastrow-Slater trial function in the VMC optimization. The introduction of an objective function allows a state-specific optimization of multiple states of the same symmetry simultaneously by imposing orthogonality between all eigenstates.

For optimization with two states, it is sufficient to apply a large enough $\lambda_{12}$ to stabilize the optimization. With more states, there are wide ranges of the various $\lambda_{IJ}$ which give consistent and stable excitation energies. Good
  values can be found at a preliminary stage with relatively few iterations.

The pure double excitations in nitroxyl [(4.32(1)~eV] and glyoxal [5.63(1)~eV] calculated here are in excellent agreement with reliable benchmarks in the literature. We also expect that our VTEs calculated at the level of DMC are reliable benchmarks for molecules and excitation types where CC methods are too costly to do with a large enough basis set. Specifically, we suspect a CC4 calculation with a larger basis set will find a tetrazine double excitation close to our result, 4.99(1)~eV, slightly lower than the CC4/6-31G* result. For cyclopentadienone, the $\mathrm{^1A_1}$ first excited state ($\mathrm{2 ^1A_1}$) is predominantly a double excitation at 5.90(1)~eV, while the second excited state is predominantly a single excitation at 6.89(1)~eV. 

The favorable computational scaling of QMC with the system size and its alignment with modern supercomputers
makes it a serious candidate for applications involving complex excited states,
such as long range charge-transfer excitations, conical
intersections, or problems where the ordering between excited states of
different character is unclear. 

\begin{acknowledgments}
We thank Pierre-Fran\c{c}ois Loos for useful discussions.
This work was supported by the European Centre of
Excellence in Exascale Computing TREX --- Targeting Real Chemical
Accuracy at the Exascale. This project has received funding from the
European Union's Horizon 2020 --- Research and Innovation program ---
under grant agreement no.~952165.
The calculations were carried out on the Dutch national supercomputer Cartesius
with the support of SURF Cooperative.
\end{acknowledgments}

\nocite{*}
\bibliography{bibliography}

\end{document}


\clearpage

\section{CASSCF Calculations}

All SA($N_{\mathrm{states}}$)-CASSCF($N_{\mathrm{e}}$,$M_{\mathrm{orb}}$) calculations are performed in GAMESS(US)~\cite{GAMESS} using BFD pseudopotentials with the corresponding aug-cc-pVDZ basis set. All geometries used are from ground-state optimizations performed at the level of CC3/aug-cc-pVTZ.~\cite{questdb,loos2019}.

All states of interest belong to the same irreducible representation as the ground state, so they have the fully-symmetric representation of the molecular geometry point-group symmetry. The point group symmetries used for nitroxyl, glyoxal, s-tetrazine, and cyclopentadienone are $\mathrm{C_s}$, $\mathrm{C_{2h}}$, $\mathrm{D_{2h}}$, and $\mathrm{C_{2v}}$, respectively.

For all molecules, we report below the sizes of the CAS used, and the total and excitation energies obtained at the CASSCF level.

\subsection{Nitroxyl}

\begin{table*}[!htb]
\caption{Summary of SA(2)-CASSCF($N_{\mathrm{e}}$,$M_{\mathrm{orb}}$) calculations on nitroxyl. The CASSCF energies ($E$) and excitation energies ($\Delta E$) are given in Hartree (H) and eV, respectively. All reference calculations are all-electron SA(2)-CASSCF calculations.}
\begin{tabular}{lrrrl}
\hline
 & \multicolumn{1}{c}{\# CSFs} & \multicolumn{1}{c}{$E(1 ^1\mathrm{A'})$} & \multicolumn{1}{c}{$E(2 ^1\mathrm{A'})$} & \multicolumn{1}{c}{$\Delta E$} \\
\hline
CAS(2,2)              &    2 &  -25.997981 &  -25.823970 & 4.74 \\
CAS(12,9)             & 1316 &  -26.127782 &  -25.955779 & 4.68 \\
\hline
CAS(12,9)/AVDZ$^a$ &      & -129.932723 & -129.760088 & 4.70 \\
CAS(12,9)/AVQZ$^a$ &      & -129.969310 & -129.797750 & 4.67 \\
\hline
\multicolumn{5}{l}{$^a$Ref.~\citenum{loos2019}} \\
\end{tabular}
\label{tab:nitroxyl-cas}
\end{table*}


\clearpage

\subsection{Glyoxal}

\begin{table*}[!htb]
\caption{Summary of SA(2)-CASSCF($N_{\mathrm{e}}$,$M_{\mathrm{orb}}$) calculations on glyoxal. The CASSCF energies ($E$) and excitation energies ($\Delta E$) are given in Hartree (H) and eV, respectively. All reference calculations are all-electron SA(2)-CASSCF calculations.}
\begin{tabular}{lrrrl}
\hline
 & \multicolumn{1}{c}{\# CSFs} & \multicolumn{1}{c}{$E(1 ^1\mathrm{A_{g}})$} & \multicolumn{1}{c}{$E(2 ^1\mathrm{A_{g}})$} & \multicolumn{1}{c}{$\Delta E$} \\
\hline
CAS(4,4)               &     8 &  -43.735960 &  -43.568786 & 4.55 \\
CAS(8,6)               &    33 &  -43.810454 &  -43.597550 & 5.79 \\
CAS(14,12)             & 42756 &  -43.922989 &  -43.705180 & 5.93 \\
\hline
CAS(8,6)/AVDZ$^a$   &       & -226.685161 & -226.471894 & 5.80 \\
CAS(8,6)/AVQZ$^a$   &       & -226.752190 & -226.540026 & 5.77 \\
CAS(14,12)/AVDZ$^a$ &       & -226.794485 & -226.576274 & 5.94 \\
CAS(14,12)/AVQZ$^a$ &       & -226.862224 & -226.644714 & 5.92 \\
\hline
\multicolumn{5}{l}{$^a$Ref.~\citenum{loos2019}} \\
\end{tabular}
\label{tab:glyoxal-cas}
\end{table*}



\subsection{Tetrazine}

\begin{table*}[!htb]
\caption{Summary of SA(2)-CASSCF($N_{\mathrm{e}}$,$M_{\mathrm{orb}}$) calculations on tetrazine. The MCSCF energies ($E$) and excitation energies ($\Delta E$) are given in Hartee (H) and eV, respectively. All reference calculations are all-electron SA(2)-CASSCF calculations.}
\begin{tabular}{lrrrl}
\hline
 & \multicolumn{1}{c}{\# CSFs} & \multicolumn{1}{c}{$E(1 ^1\mathrm{A_{1g}})$} & \multicolumn{1}{c}{$E(2 ^1\mathrm{A_{1g}})$} & \multicolumn{1}{c}{$\Delta E$} \\
\hline
CAS(4,4)                     &   6 &  -51.028815 &  -50.806971 & 6.04 \\
CAS(8,6)                     &  21 &  -51.079408 &  -50.821626 & 7.02 \\
CAS(12,8)                    &  56 &  -51.079992 &  -50.846778 & 6.35 \\
CAS(14,10)                   & 670 &  -51.130154 &  -50.930008 & 5.45 \\
\hline
CAS(14,10)/AVDZ$^a$ &     & -294.727767 & -294.526292 & 5.48 \\
CAS(14,10)/AVQZ$^a$ &     & -294.803516 & -294.604540 & 5.41 \\
\hline
\multicolumn{5}{l}{$^a$Ref.~\citenum{loos2019}} \\
\end{tabular}
\label{tab:tetrazine-cas}
\end{table*}


\clearpage

\subsection{Cyclopentadienone}

\begin{table*}[!htb]
\caption{Summary of SA($N_{\mathrm{state}}$)-CASSCF($N_{\mathrm{e}}$, $M_{\mathrm{orb}}$) calculations on cyclopentadienone. The CASSCF energies ($E$) and excitation energies ($\Delta E$) are given in Hartree (H) and eV, respectively.}
\resizebox{\columnwidth}{!}{
\begin{tabular}{lrllllll}
\hline
 & \multicolumn{1}{c}{\# CSFs} & \multicolumn{1}{c}{$E(1 ^1\mathrm{A_{1}})$} & \multicolumn{1}{c}{$E(2 ^1\mathrm{A_{1}})$} & \multicolumn{1}{c}{$E(3 ^1\mathrm{A_{1}})$} & \multicolumn{1}{c}{$\Delta E_{12}$} & \multicolumn{1}{c}{$\Delta E_{13}$} & \multicolumn{1}{c}{$\Delta E_{23}$} \\
\hline
\multicolumn{8}{l}{SA(3)} \\
\cline{1-1}
CAS(6,6)                     &   95 &  -45.749640 &  -45.519741 & -45.437557 & 6.26 & 8.49 & 2.24 \\
CAS(8,7)                     &  152 &  -45.749865 &  -45.519864 & -45.437841 & 6.26 & 8.49 & 2.23 \\
CAS(10,8)                    &  326 &  -45.752040 &  -45.521488 & -45.438852 & 6.27 & 8.52 & 2.25 \\[.5ex]
\multicolumn{8}{l}{SA(4)} \\
\cline{1-1}
CAS(6,6)                     &   95 &  -45.747679 &  -45.520726 & -45.434884 & 6.18 & 8.51 & 2.34 \\
CAS(8,7)                     &  152 &  -45.747963 &  -45.520831 & -45.435133 & 6.18 & 8.51 & 2.33 \\
CAS(10,8)                    &  326 &  -45.750131 &  -45.522663 & -45.435928 & 6.19 & 8.55 & 2.36 \\[.5ex]
\multicolumn{8}{l}{SA(5)} \\
\cline{1-1}
CAS(6,6)                     &   95 &  -45.744381 &  -45.518430 & -45.432765 & 6.15 & 8.48 & 2.33 \\
CAS(8,7)                     &  152 &  -45.747246 &  -45.518198 & -45.435403 & 6.23 & 8.49 & 2.25 \\
CAS(10,8)                    &  326 &  -45.748150 &  -45.519019 & -45.436202 & 6.23 & 8.49 & 2.25 \\
\hline
\end{tabular}
}
\label{tab:cyclopentadienone-cas}
\end{table*}

\clearpage

\section{CIPSI calculations}\label{sec:cipsi-dets}

\subsection{PT2 and Variance Matching}\label{sec:cipsi-dets-matching}
    
At each iteration in the CIPSI algorithm, a new set of determinants is added to the determinant space, shared by all states. In Quantum Package~\cite{QP,QP2} (QP), one controls how many determinants are added at each iteration via the selection factor, $f$, given by,
\begin{equation}\label{eq:sf}
 N_{\alpha}=\mathrm{INT}(\sqrt{N_{\mathrm{states}}}\times N_{\mathrm{det}}\times f),
\end{equation}
where $N_{\alpha}$ is the number of new determinants to add to the expansion and $N_{\mathrm{det}}$ is the number determinants currently in the expansion. 

The wave functions of the different states are said to be `matched' when they have similar PT2 energies ($E_{\mathrm{PT2}}$) and variances ($\sigma^2$), which quantify the quality of the wave function. In QP, the matching is controlled at each iteration by calculating selection weights on each state, $W_I$, which then bias the selection of determinants for the improvement of a given state. An external
determinant $\ket{\alpha}$ is selected based on its weighted PT2 energy contribution 
obtained via the Epstein-Nesbet partitioning of the Hamiltonian,
\begin{equation}
e_{\alpha} = \sum_{I=1}^{N_{\rm states}} W_I\delta E_{\alpha,I}^{(2)},
\end{equation}
where
\begin{equation}
\delta E_{\alpha,I}^{(2)} = \frac{|\langle\alpha|H|\Psi_I^{\rm CIPSI}\rangle|^2}{\langle\Psi_I^{\rm CIPSI}|H|\Psi_I^{\rm CIPSI}\rangle - \langle\alpha|H|\alpha\rangle}\,,
\label{eq:en-pt2}
\end{equation}
or based on its weighted contribution to the variance of the states
\begin{equation}
\sigma^2_{\alpha} = \sum_{I=1}^{N_{\rm states}} W_I\delta \sigma^2_{\alpha,I},
\end{equation}
where
\begin{equation}
\delta \sigma^2_{\alpha,I} = |\langle\alpha|H|\Psi_I^{\rm CIPSI}\rangle|^2,
\label{eq:variance}
\end{equation}
and $\Psi_{I}^{\rm CIPSI}$ is the current normalized CIPSI wave function for state $I$. 
At a given iteration, the larger the selection weight for a state, the more likely determinants will be added to the expansion which benefit that state. The goal is to keep the quality of the wave functions similar during the expansion.

The simplest way to calculate $W_I$ is by using the largest expansion coefficient in the current wave function for each state, $^ic_I^{\mathrm{max}}$, where $i$ is the current iteration. The iteration index will be removed until a distinction between iterations must be made. The unnormalized weight is given by, $\tilde{w}_I^{\mathrm{c_{max}}}=1/|c_I^{\mathrm{max}}|^2$. The normalized weights are given by
\begin{equation}\label{eq:cmax}
 w_I^{\mathrm{c_{max}}}=\frac{\tilde{w}_I^{\mathrm{c_{max}}}}{\sum_I^{N_{\mathrm{states}}}\tilde{w}_I^{\mathrm{c_{max}}}}.
\end{equation}
Two different approaches are used to calculate $W_I$, both of which include the weights in Eq.~\ref{eq:cmax}.

In the first approach (i), additional input weights, $w_I^{\mathrm{inp}}$, (which sum to unity) are chosen by hand and they remain fixed throughout the calculation~\cite{dash2021}. These are the same weights denoted as $w_{I}^{\mathrm{SA}}$ in Ref.~\cite{dash2021}, but we wish to emphasize here that these weights are provided as input. The input weights enter the selection weights at each iteration as
\begin{equation}\label{eq:algo0}
\tilde{W}_I=(w_I^{\mathrm{c_{max}}})\times(w_I^{\mathrm{inp}}),
\end{equation}
where a tilde symbol indicates that the weight is not normalized.

In the second approach (ii), weights are constructed from the PT2 energies, $w_I^{\mathrm{PT2}}$, and variance, $w_I^{\mathrm{\sigma^2}}$, and are used to calculate the selection weights. These two unnormalized weights are calculated according to
\begin{equation}\label{eq:pt2}
 ^i\tilde{w}_I^{\mathrm{PT2}}=(^{i-1}w_I^{\mathrm{PT2}})\times(^iE^{\mathrm{PT2}}_I),
\end{equation}
\begin{equation}\label{eq:var}
 ^i\tilde{w}_I^{\mathrm{\sigma^2}}=(^{i-1}w_I^{\mathrm{\sigma^2}}).\times(^i\sigma^2_I)
\end{equation}
Note that the unnormalized weights at the present iteration, $i$, are calculated using the normalized weights from the previous iteration, which are set to unity at the start of the calculation. These weights do not affect the selection weight until 100 determinants have been reached. After the weights in Eqs.~\ref{eq:pt2}~and~\ref{eq:var} are normalized in the same way as in Eq.~\ref{eq:cmax}, the unnormalized selection weights for approach (ii) are calculated as
\begin{equation}
\tilde{W}_I=(w_I^{\mathrm{c_{max}}})\times \sqrt{(w_I^{\mathrm{PT2}})\times(w_I^{\mathrm{\sigma^2}})}.
\end{equation}
Therefore, the selection weight is larger for a state with smaller $c_I^\mathrm{max}$, but larger $E^{\mathrm{PT2}}_I$ and $\sigma^2_I$, maintaining wave functions of similar quality.

Approach (i) is used for all molecules, while (ii) is only utilized for cyclopentadienone. Using different input weights in approach (i) will lead to a different evolution of the determinant expansion, and a slightly different number of determinants added at each CIPSI iteration. Therefore, to compare expansions with different input weights in Section~\ref{sec:cipsi-data}, expansions with similar number of determinants are provided as best as possible. 
The optimal CIPSI input weights, $w_I^{\mathrm{inp}}$, for different systems together with the starting orbitals and expansion are given in the next Section.


\subsection{Starting Wave Functions}

In a multi-state CIPSI calculation it is important to make sure the correct excited state(s) are found early in the CIPSI expansion. In general, performing a CI including singles and doubles (CISD) calculation ensures the lowest energy states in a given symmetry are found. The CISD can be performed iteratively in QP and stopped after reaching a predetermined number of determinants, $N_{\mathrm{det}}^{\mathrm{max}}$. 

For nitroxyl, glyoxal, and tetrazine, the excitations of interest are dominated by closed-shell doubly-excited determinants, so it is more appropriate to perform a seniority-zero, doubly occupied CI (DOCI) to build the initial CIPSI wave function. 
This determinant expansion is then truncated to include the determinants with largest expansion coefficients (between 2 and 51 determinants). 
This initial wave function ensures the double excitation of interest is found immediately in the CIPSI calculation. 
Starting from this wave function, an unrestricted CIPSI is performed up to about $10^4$ determinants. 


For cyclopentadienone, the excited-states of interest have a large mix of singly- and doubly-excited determinants so an iterative CISD up to $\sim$$10^3$ determinants is performed and then truncated to $\sim$$100$ determinants to initialize the wave function. 

It is found, in general, that as long as the dominant determinant is included in the starting wave function the CIPSI expansion will follow a similar evolution. This is made clear by Figure~\ref{fig:cipsi-E-cas-vs-no} below. The CIPSI performed on glyoxal starting from a DOCI truncated to 2 determinants (DOCI[HF+1]) versus 51 determinants (DOCI[HF+50]) made little difference in the evolution of the energies beyond 50 determinants. Data on all starting CIPSI wave functions are provided in Table~\ref{tab:cipsi-orb-wf}.

\begin{table*}[!htb]
\caption{Summary of CIPSI starting wave functions (WF), orbitals, and optimal weights ($w_I^{\mathrm{inp}}$) providing the best PT2 and variance matching. For the NO orbitals, we denote with \emph{method}(\emph{HF+\# dets after truncation})-\emph{\# dets in CIPSI} the wave function from which the NO orbitals are calculated: for instance, DOCI(HF+1)-10000 means that the NO orbitals are calculated from a first CIPSI expansion to about 10000 determinants, starting from a DOCI truncated to 2 determinants. 
Note that the matching algorithm (ii) described in the previous Section is used for cyclopentadienone in the NO basis.}
\resizebox{\columnwidth}{!}{
\begin{tabular}{lllcllc}
\hline
 & \multicolumn{3}{c}{CAS} & \multicolumn{3}{c}{NO} \\
 \cline{2-4}
 \cline{5-7}
\multicolumn{1}{c}{Molecule} & \multicolumn{1}{c}{Orbitals} & \multicolumn{1}{c}{WF} & \multicolumn{1}{c}{$w_I^{\mathrm{inp}}$} & \multicolumn{1}{c}{Orbitals} & \multicolumn{1}{c}{WF} & \multicolumn{1}{c}{$w_I^{\mathrm{inp}}$}  \\
\hline
Nitroxyl               & SA(2)-CASSCF(12,9) & DOCI(HF+1) & \multicolumn{1}{c}{0.525$-$0.475}  & DOCI(HF+1)-10000  & DOCI(HF+1) & 0.525$-$0.475 \\
Glyoxal                & SA(2)-CASSCF(14,12) & DOCI(HF+1) & \multicolumn{1}{c}{0.50$-$0.50} & DOCI(HF+50)-10000 & DOCI(HF+50) & 0.45$-$0.55 \\
Tetrazine              & SA(2)-CASSCF(14,10) & DOCI(HF+1) & \multicolumn{1}{c}{0.45$-$0.55} & DOCI(HF+50)-10000 & DOCI(HF+50) & 0.45$-$0.55 \\
Cyclopentadienone      & SA(3)-CASSCF(6,6) & CISD(HF+97) & \multicolumn{1}{c}{0.27$-$0.35$-$0.38} & CISD(HF+97)-20000 & CISD(HF+101) & \multicolumn{1}{c}{$-$} \\
\hline
\end{tabular}
}
\label{tab:cipsi-orb-wf}
\end{table*}


\clearpage

\subsection{Molecular Orbital Basis}

\begin{figure}[htb]
\begin{center}
\includegraphics[scale=0.45]{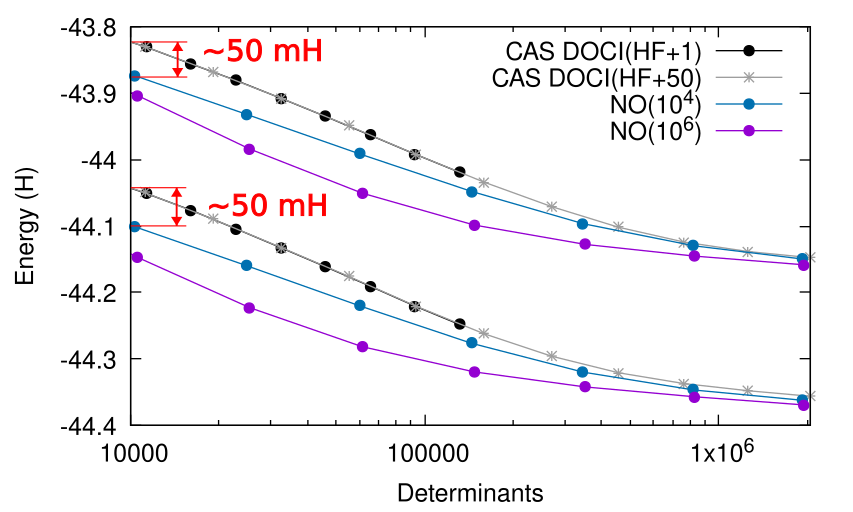}
\caption{\label{fig:cipsi-E-cas-vs-no} Glyoxal ground- and excited-state CIPSI energies using different starting wave functions and MO basis. The two CIPSI expansions performed in the CAS basis show that different starting wave functions lead to the same evolution of the energies. The two CIPSI expansions performed in the NO basis both start from a DOCI(HF+50) wave function, but one set of NOs were generated from a previous CIPSI wave function with $10^4$ determinants and the other with $10^6$. The same input weights ($w_1^{\mathrm{inp}}$~$w_2^{\mathrm{inp}}$~:~0.5~0.5) and basis set are used in all calculations.}
\end{center}
\end{figure}

The CIPSI expansions are performed in a molecular orbital (MO) basis built from either the CASSCF optimized orbitals of a SA($N_{\mathrm{states}}$)-CASSCF($N_{\mathrm{e}}$,$M_{\mathrm{orb}}$) calculation (referred to as CAS) or the natural orbitals (NO) built from a CIPSI expansion performed in the CAS basis.


The choice of MO basis (CAS versus NO) has a noticeable effect on the energies in the CIPSI expansion. Figure~\ref{fig:cipsi-E-cas-vs-no} illustrates how the ground- and excited-state energies of glyoxal are affected when choosing different starting wave functions and size of the CIPSI used to build the NOs. The energies of the ground- and excited-states all tend towards the same energy as the expansion reaches $10^6$ determinants, but the energies around $10^4$ determinants are noticeably different for different MO basis. Using NOs from a $10^4$ and a $10^6$ determinant expansion lowers the energy of both states by $\sim$50~mH and $\sim$70~mH, respectively, compared to the CAS basis. This suggests a better quality wave function is obtained in the NO basis for the same size determinant expansions. This is also expected to lead to a faster convergence of the VMC/DMC VTEs at smaller determinant expansion. Details on the MO basis for all CIPSI calculations are provided in Table~\ref{tab:cipsi-orb-wf}.


\section{CIPSI Data}\label{sec:cipsi-data}

In the following, relevant data from CIPSI calculations are provided, including CIPSI energies ($E_{\mathrm{CIPSI}}$), PT2 energies ($E_{\mathrm{PT2}}$), and variances ($\sigma^2$) for ground and excited states.  The differences are also provided and are always calculated as $\Delta X =X_{\mathrm{excited}}-X_{\mathrm{ground}}$. The Tables in this Section have some values in boldface which correspond the CIPSI wave functions that are used as VMC and DMC trial wave functions either in the SI or in the main text. 

\clearpage

\subsection{Nitroxyl}

\begin{figure}[htb]
\begin{center}
\includegraphics[scale=0.363]{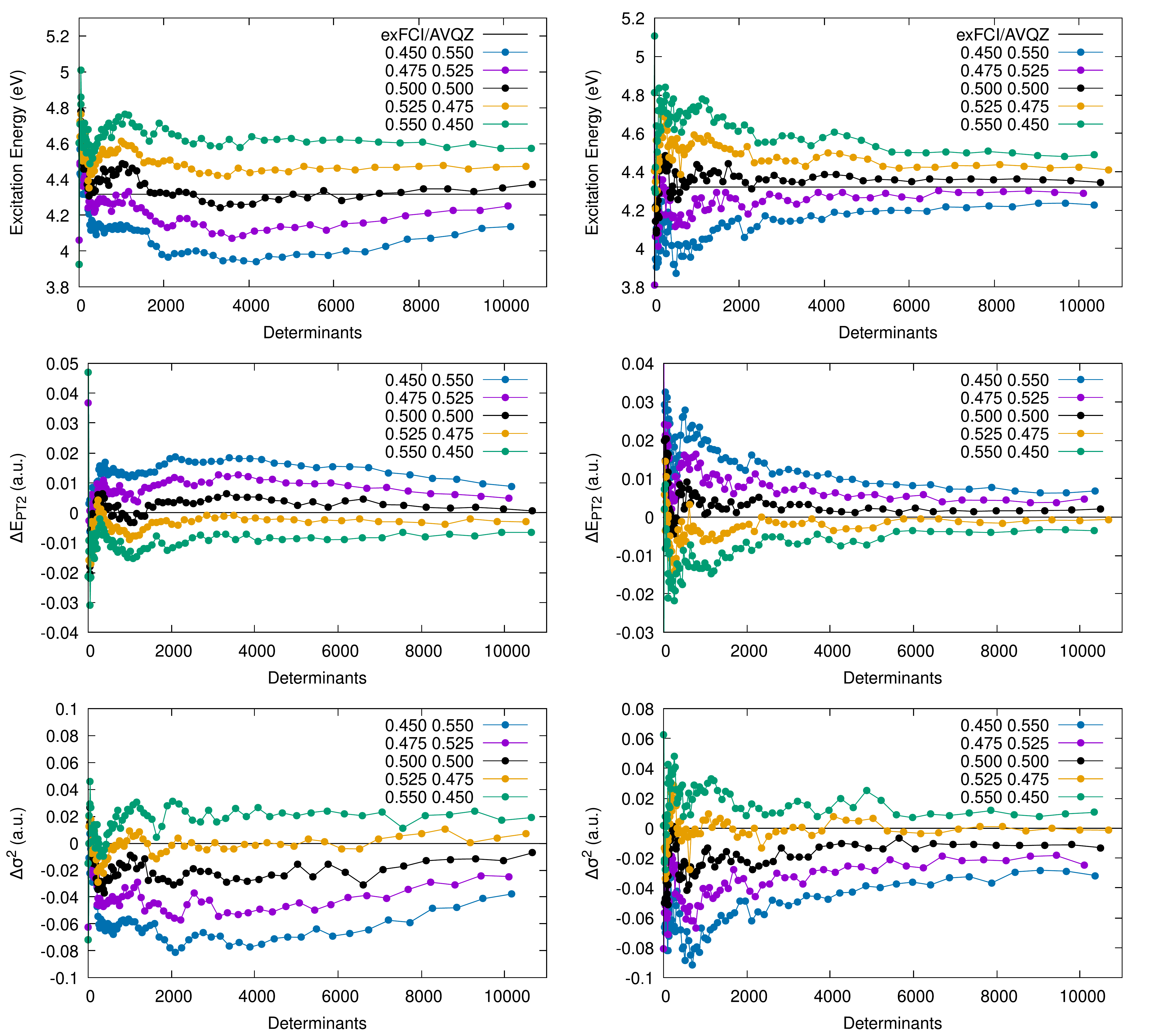}
\caption{\label{fig:nitroxyl-cipsi} Nitroxyl CIPSI energies, PT2 energies, and variances in the CAS (left) and NO (right) basis, using different input weights ($w_1^{\mathrm{inp}}$ $w_2^{\mathrm{inp}}$).}
\end{center}
\end{figure}

\begin{table*}[!htb]
\caption{Nitroxyl CIPSI energies ($E_{\mathrm{CIPSI}}$), PT2 energies ($E_{\mathrm{PT2}}$), and variances ($\sigma^2$) of ground- and excited-state CIPSI expansions for different input weights, $w_I^{\mathrm{inp}}$. The CIPSI excitation energy, $\Delta E$, is given in eV, the remaining values are in a.u. The optimal weights ($w_1^{\mathrm{inp}}-w_2^{\mathrm{inp}}$) for both CAS and NO orbitals are 0.525$-$0.475.}
\resizebox{\columnwidth}{!}{
\begin{tabular}{lrrrrrrrrr}
\hline
        & \multicolumn{3}{c}{$E_{\mathrm{CIPSI}}$} & \multicolumn{3}{c}{$E_{\mathrm{PT2}}$} & \multicolumn{3}{c}{$\sigma^2$} \\
\cline{2-4}
\cline{5-7}
\cline{8-10}
\# det. & \multicolumn{1}{c}{$1 ^1\mathrm{A'}$} & \multicolumn{1}{c}{$2 ^1\mathrm{A'}$} & \multicolumn{1}{c}{$\Delta E$} & \multicolumn{1}{c}{$1 ^1\mathrm{A'}$} & \multicolumn{1}{c}{$2 ^1\mathrm{A'}$} & \multicolumn{1}{c}{$\Delta E_{\mathrm{PT2}}$} & \multicolumn{1}{c}{$1 ^1\mathrm{A'}$} & \multicolumn{1}{c}{$2 ^1\mathrm{A'}$} & \multicolumn{1}{c}{$\Delta\sigma^2$} \\
\hline
\multicolumn{10}{c}{CAS 0.50$-$0.50} \\
2     & -25.9894 & -25.8203 & 4.602 & -0.5649 & -0.5861 & -0.0212 & 1.5944 & 1.5793 & -0.0151 \\
59    & -26.0943 & -25.9193 & 4.762 & -0.3468 & -0.3638 & -0.0170 & 1.2972 & 1.3152 &  0.0180 \\
129   & -26.1106 & -25.9417 & 4.596 & -0.3219 & -0.3300 & -0.0081 & 1.2517 & 1.2527 &  0.0010 \\
\bf{363}   & -26.1292 & -25.9708 & \bf{4.310} & -0.2968 & -0.2901 &  \bf{0.0067} & 1.1848 & 1.1484 & \bf{-0.0364} \\
600   & -26.1476 & -25.9841 & 4.448 & -0.2733 & -0.2742 & -0.0009 & 1.1203 & 1.1019 & -0.0184 \\
1268  & -26.1756 & -26.0124 & 4.441 & -0.2394 & -0.2401 & -0.0007 & 1.0083 & 0.9952 & -0.0131 \\
\bf{1627}  & -26.1849 & -26.0241 & \bf{4.374} & -0.2279 & -0.2258 &  \bf{0.0021} & 0.9723 & 0.9484 & \bf{-0.0239} \\
2686  & -26.2087 & -26.0498 & 4.324 & -0.2006 & -0.1971 &  0.0036 & 0.8746 & 0.8542 & -0.0204 \\
\bf{3447}  & -26.2206 & -26.0643 & \bf{4.252} & -0.1867 & -0.1813 &  \bf{0.0053} & 0.8236 & 0.7975 & \bf{-0.0262} \\
\bf{10660} & -26.2846 & -26.1238 & \bf{4.374} & -0.1205 & -0.1198 &  \bf{0.0006} & 0.5826 & 0.5754 & \bf{-0.0072} \\
\hline
\multicolumn{10}{c}{CAS 0.525$-$0.475} \\
2     & -25.9894 & -25.8203 & 4.602 & -0.5649 & -0.5861 & -0.0212 & 1.5944 & 1.5793 & -0.0151 \\
59    & -26.0943 & -25.9193 & 4.762 & -0.3468 & -0.3638 & -0.0170 & 1.2972 & 1.3152 &  0.0180 \\
136   & -26.1129 & -25.9412 & 4.672 & -0.3181 & -0.3305 & -0.0124 & 1.2464 & 1.2543 &  0.0079 \\
\bf{322}   & -26.1278 & -25.9652 & \bf{4.424} & -0.2986 & -0.2971 &  \bf{0.0015} & 1.1891 & 1.1678 & \bf{-0.0213} \\
600   & -26.1494 & -25.9823 & 4.547 & -0.2710 & -0.2762 & -0.0053 & 1.1122 & 1.1071 & -0.0050 \\
1235  & -26.1773 & -26.0086 & 4.590 & -0.2370 & -0.2444 & -0.0074 & 0.9993 & 1.0080 & 0.0086 \\
\bf{1569}  & -26.1853 & -26.0204 & \bf{4.487} & -0.2273 & -0.2302 &  \bf{-0.0029} & 0.9701 & 0.9576 & \bf{-0.0126} \\
2657  & -26.2108 & -26.0469 & 4.460 & -0.1982 & -0.2002 & -0.0020 & 0.8662 & 0.8628 & -0.0035 \\
\bf{3751}  & -26.2287 & -26.0647 & \bf{4.464} & -0.1786 & -0.1809 &  \bf{-0.0022} & 0.7961 & 0.7951 & \bf{-0.0010} \\
\bf{10516} & -26.2856 & -26.1213 & \bf{4.473} & -0.1196 & -0.1227 &  \bf{-0.0031} & 0.5788 & 0.5858 & \bf{0.0070} \\
\hline
\multicolumn{10}{c}{NO 0.525$-$0.475} \\
2     & -25.9964 & -25.8197 & 4.810 & -0.5825 & -0.6169 & -0.0344 & 1.5820 & 1.5837 & 0.0016 \\
64    & -26.1015 & -25.9388 & 4.426 & -0.3711 & -0.3672 &  0.0040 & 1.2827 & 1.2704 &  -0.0123 \\
133   & -26.1257 & -25.9575 & 4.576 & -0.3328 & -0.3397 & -0.0068 & 1.1964 & 1.2034 &  0.0070 \\
\bf{321}   & -26.1578 & -25.9913 & \bf{4.529} & -0.2879 & -0.2940 &  \bf{-0.0062} & 1.0784 & 1.0786 & \bf{0.0002} \\
617   & -26.1846 & -26.0235 & 4.386 & -0.2522 & -0.2489 & 0.0033 & 0.9877 &  0.9603 & -0.0274 \\
1257  & -26.2283 & -26.0597 & 4.588 & -0.1966 & -0.2027 & -0.0061 & 0.8156 & 0.8196 & 0.0040 \\
\bf{1573}  & -26.2411 & -26.0742 & \bf{4.542} & -0.1812 & -0.1855 &  \bf{-0.0042} & 0.7655 & 0.7683 & \bf{0.0027} \\
2699  & -26.2704 & -26.1066 & 4.455 & -0.1470 & -0.1484 & -0.0014 & 0.6549 & 0.6524 & -0.0026 \\
\bf{2900}  & -26.2743 & -26.1102 & \bf{4.466} & -0.1425 & -0.1442 &  \bf{-0.0017} & 0.6395 & 0.6379 & \bf{-0.0016} \\
\bf{10690} & -26.3292 & -26.1671 & \bf{4.410} & -0.0813 & -0.0820 &  \bf{-0.0006} & 0.4124 & 0.4111 & \bf{-0.0013} \\

\hline
\end{tabular}
}
\label{tab:nitroxyl-cipsi}
\end{table*}






\clearpage

\subsection{Glyoxal}


\begin{figure}[htb]
\begin{center}
\includegraphics[scale=0.363]{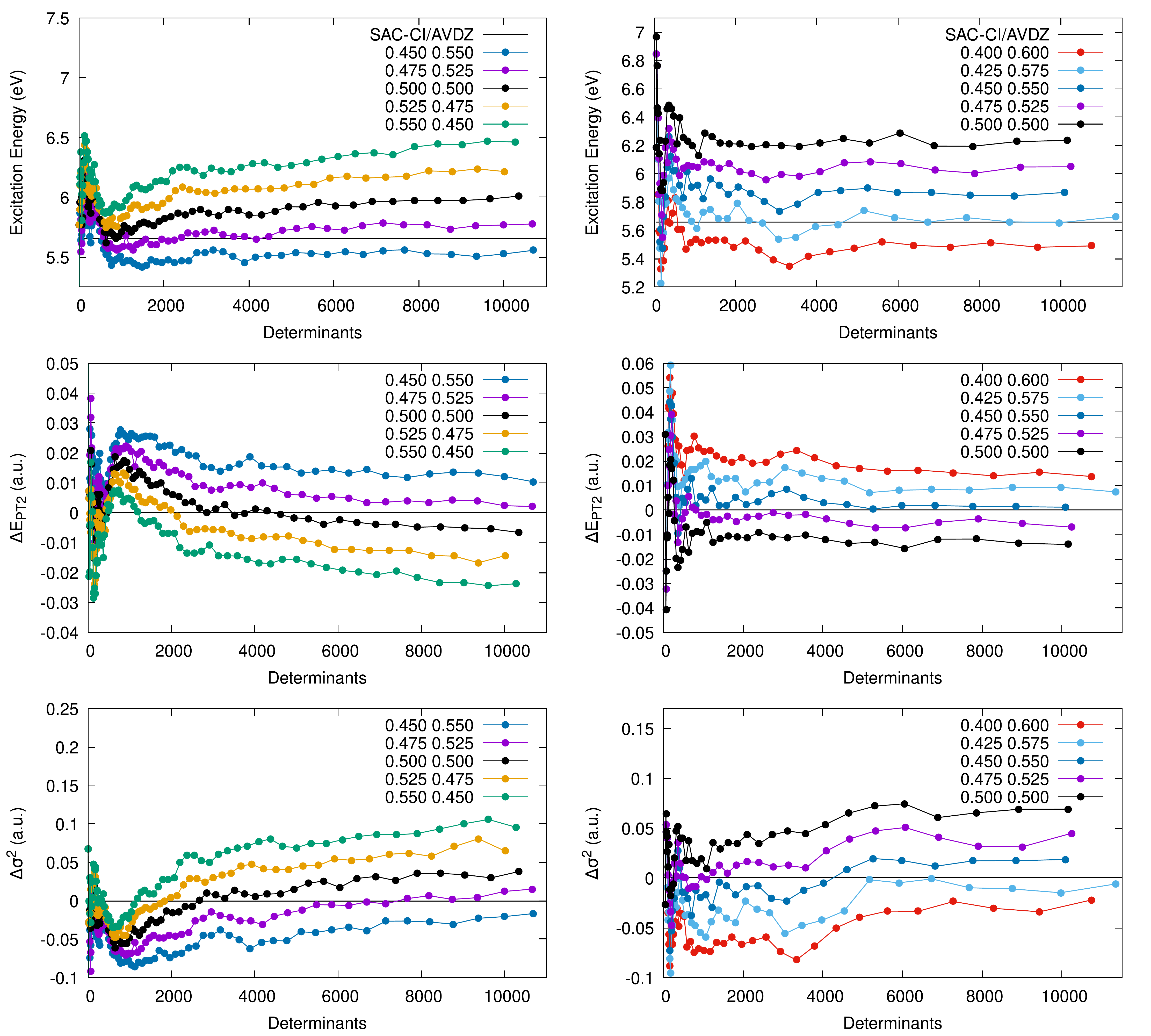}
\caption{\label{fig:glyoxal-cipsi} Glyoxal CIPSI energies, PT2 energies, and variances in the CAS (left) and NO (right) basis, using different input weights ($w_1^{\mathrm{inp}}$ $w_2^{\mathrm{inp}}$).}
\end{center}
\end{figure}

\begin{table*}[!htb]
\caption{Glyoxal CIPSI energies ($E_{\mathrm{CIPSI}}$), PT2 energies ($E_{\mathrm{PT2}}$), and variances ($\sigma^2$) of ground- and excited-state CIPSI expansions using different input weights, $w_I^{\mathrm{inp}}$. The CIPSI excitation energy, $\Delta E$, is given in eV, the remaining values are in a.u. The optimal weights ($w_1^{\mathrm{inp}}-w_2^{\mathrm{inp}}$) for CAS are 0.50$-$0.50 and for NO are 0.45$-$0.55.}
\resizebox{\columnwidth}{!}{
\begin{tabular}{lrrrrrrrrr}
\hline
        & \multicolumn{3}{c}{$E_{\mathrm{CIPSI}}$} & \multicolumn{3}{c}{$E_{\mathrm{PT2}}$} & \multicolumn{3}{c}{$\sigma^2$} \\
\cline{2-4}
\cline{5-7}
\cline{8-10}
\# det. & \multicolumn{1}{c}{$1 ^1\mathrm{A_g}$} & \multicolumn{1}{c}{$2 ^1\mathrm{A_g}$} & \multicolumn{1}{c}{$\Delta E$} & \multicolumn{1}{c}{$1 ^1\mathrm{A_g}$} & \multicolumn{1}{c}{$2 ^1\mathrm{A_g}$} & \multicolumn{1}{c}{$\Delta E_{\mathrm{PT2}}$} & \multicolumn{1}{c}{$1 ^1\mathrm{A_g}$} & \multicolumn{1}{c}{$2 ^1\mathrm{A_g}$} & \multicolumn{1}{c}{$\Delta\sigma^2$} \\
\hline
\multicolumn{10}{c}{CAS 0.475$-$0.525} \\
2     & -43.7144 & -43.4030 & 8.473 & -0.9159 & -1.2042 & -0.2883 & 2.8205 & 2.8883 &  0.0678 \\
57    & -43.8122 & -43.6084 & 5.543 & -0.6909 & -0.6526 &  0.0383 & 2.5581 & 2.4665 & -0.0916 \\
127   & -43.8488 & -43.6247 & 6.097 & -0.6275 & -0.6285 & -0.0010 & 2.4248 & 2.4309 &  0.0061 \\
372   & -43.8848 & -43.6707 & 5.827 & -0.5696 & -0.5626 &  0.0070 & 2.3218 & 2.2899 & -0.0319 \\
582   & -43.8995 & -43.6910 & 5.675 & -0.5494 & -0.5342 &  0.0151 & 2.2827 & 2.2291 & -0.0536 \\
1281  & -43.9266 & -43.7224 & 5.557 & -0.5152 & -0.4947 &  0.0205 & 2.1864 & 2.1236 & -0.0628 \\
1582  & -43.9369 & -43.7293 & 5.649 & -0.5025 & -0.4873 &  0.0152 & 2.1458 & 2.1012 & -0.0446 \\
2588  & -43.9607 & -43.7510 & 5.706 & -0.4730 & -0.4642 &  0.0088 & 2.0476 & 2.0210 & -0.0266 \\
3401  & -43.9746 & -43.7655 & 5.689 & -0.4562 & -0.4472 &  0.0090 & 1.9826 & 1.9609 & -0.0217 \\
\bf{10660} & -44.0434 & -43.8311 & \bf{5.777} & -0.3792 & -0.3771 &  \bf{0.0021} & 1.7025 & 1.7178 &  \bf{0.0153} \\
\hline
\multicolumn{10}{c}{CAS 0.50$-$0.50} \\
2     & -43.7144 & -43.4030 & 8.473 & -0.9159 & -1.2042 & -0.2882 & 2.8205 & 2.8883 &  0.0678 \\
54    & -43.8122 & -43.6033 & 5.683 & -0.6909 & -0.6616 &  0.0294 & 2.5582 & 2.4928 & -0.0654 \\
123   & -43.8500 & -43.6220 & 6.204 & -0.6247 & -0.6328 & -0.0081 & 2.4252 & 2.4405 &  0.0153 \\
\bf{358}   & -43.8891 & -43.6649 & \bf{6.101} & -0.5636 & -0.5702 & \bf{-0.0065} & 2.3101 & 2.3044 & \bf{-0.0057} \\
596   & -43.9016 & -43.6907 & 5.740 & -0.5474 & -0.5351 &  0.0122 & 2.2791 & 2.2337 & -0.0454 \\
1270  & -43.9300 & -43.7186 & 5.752 & -0.5116 & -0.4992 &  0.0124 & 2.1758 & 2.1341 & -0.0416 \\
\bf{1632}  & -43.9404 & -43.7285 & \bf{5.766} & -0.4980 & -0.4878 &  \bf{0.0102} & 2.1275 & 2.1052 & \bf{-0.0223} \\
2687  & -43.9673 & -43.7496 & 5.923 & -0.4649 & -0.4649 & -0.0000 & 2.0154 & 2.0196 &  0.0042 \\
\bf{3446}  & -43.9792 & -43.7626 & \bf{5.894} & -0.4509 & -0.4511 & \bf{-0.0002} & 1.9632 & 1.9766 &  \bf{0.0134} \\
\bf{10342} & -44.0459 & -43.8251 & \bf{6.010} & -0.3768 & -0.3834 & \bf{-0.0066} & 1.6973 & 1.7355 &  \bf{0.0382} \\
\hline
\multicolumn{10}{c}{NO 0.45$-$0.55} \\
51    & -43.7834 & -43.5561 & 6.184 & -0.8056 & -0.7745 &  0.0311 & 2.6509 & 2.6242 & -0.0267 \\
128   & -43.8355 & -43.6183 & 5.911 & -0.6804 & -0.6616 &  0.0188 & 2.4840 & 2.4651 & -0.0189 \\
349   & -43.8907 & -43.6684 & 6.048 & -0.5890 & -0.5845 &  0.0046 & 2.3171 & 2.3142 & -0.0029 \\
595   & -43.9187 & -43.6980 & 6.005 & -0.5457 & -0.5416 &  0.0041 & 2.2218 & 2.2100 & -0.0119 \\
1276  & -43.9594 & -43.7432 & 5.884 & -0.4892 & -0.4831 &  0.0061 & 2.0775 & 2.0565 & -0.0210 \\
\bf{3749}  & -44.0262 & -43.8141 & \bf{5.772} & -0.4077 & -0.4010 &  \bf{0.0067} & 1.8119 & 1.7876 & \bf{-0.0243} \\
\bf{5046}  & -44.0485 & -43.8324 & \bf{5.879} & -0.3823 & -0.3812 &  \bf{0.0011} & 1.7075 & 1.7240 & \bf{0.0164} \\
\bf{7040}  & -44.0701 & -43.8549 & \bf{5.856} & -0.3586 & -0.3567 &  \bf{0.0020} & 1.6296 & 1.6400 & \bf{0.0104} \\
\bf{8642}  & -44.0842 & -43.8682 & \bf{5.879} & -0.3446 & -0.3451 & \bf{-0.0005} & 1.5836 & 1.6106 &  \bf{0.0270} \\
\bf{10010} & -44.0923 & -43.8785 & \bf{5.817} & -0.3357 & -0.3338 & \bf{0.0019} & 1.5494 & 1.5652 &  \bf{0.0158} \\
\bf{12132} & -44.1065 & -43.8913 & \bf{5.858} & -0.3208 & -0.3201 & \bf{0.0007} & 1.5020 & 1.5155 &  \bf{0.0134} \\
\bf{15349} & -44.1209 & -43.9080 & \bf{5.793} & -0.3061 & -0.3046 & \bf{0.0015} & 1.4541 & 1.4680 &  \bf{0.0139} \\

\hline
\end{tabular}
}
\label{tab:glyoxal-cipsi}
\end{table*}





\clearpage

\subsection{Tetrazine}




\begin{figure}[htb]
\begin{center}
\includegraphics[scale=0.363]{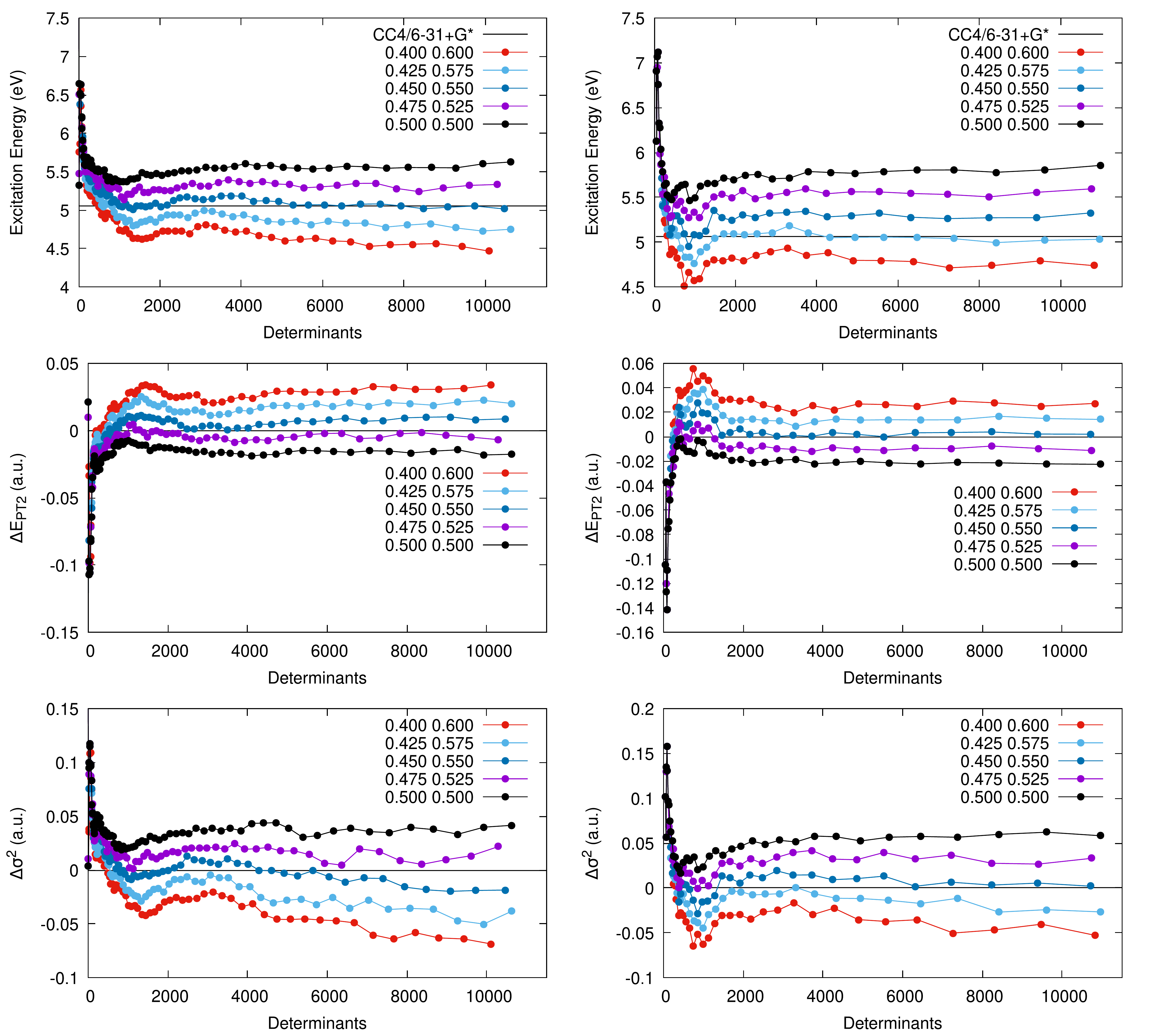}
\caption{\label{fig:tetrazine-cipsi} Tetrazine CIPSI energies, PT2 energies, and variances in the CAS (left) and NO (right) basis, using different input weights ($w_1^{\mathrm{inp}}$ $w_2^{\mathrm{inp}}$).}
\end{center}
\end{figure}

\begin{table*}[!htb]
\caption{Tetrazine CIPSI energies ($E_{\mathrm{CIPSI}}$), PT2 energies ($E_{\mathrm{PT2}}$), and variances ($\sigma^2$) of ground- and excited-state CIPSI expansions using different input weights, $w_I^{\mathrm{inp}}$. The CIPSI excitation energy, $\Delta E$, is given in eV, the remaining values are in a.u. The optimal weights ($w_1^{\mathrm{inp}}-w_2^{\mathrm{inp}}$) for both CAS and NO are 0.45$-$0.55.}
\resizebox{\columnwidth}{!}{
\begin{tabular}{lrrrrrrrrr}
\hline
        & \multicolumn{3}{c}{$E_{\mathrm{CIPSI}}$} & \multicolumn{3}{c}{$E_{\mathrm{PT2}}$} & \multicolumn{3}{c}{$\sigma^2$} \\
\cline{2-4}
\cline{5-7}
\cline{8-10}
\# det. & \multicolumn{1}{c}{$1 ^1\mathrm{A_{1g}}$} & \multicolumn{1}{c}{$2 ^1\mathrm{A_{1g}}$} & \multicolumn{1}{c}{$\Delta E$} & \multicolumn{1}{c}{$1 ^1\mathrm{A_{1g}}$} & \multicolumn{1}{c}{$2 ^1\mathrm{A_{1g}}$} & \multicolumn{1}{c}{$\Delta E_{\mathrm{PT2}}$} & \multicolumn{1}{c}{$1 ^1\mathrm{A_{1g}}$} & \multicolumn{1}{c}{$2 ^1\mathrm{A_{1g}}$} & \multicolumn{1}{c}{$\Delta\sigma^2$} \\
\hline
\multicolumn{10}{c}{CAS 0.45$-$0.55} \\
2     & -51.0258 & -50.7256 & 8.168 & -1.2075 & -1.5672 & -0.3597 & 3.4424 & 3.6206 &  0.1783 \\
61    & -51.1111 & -50.8776 & 6.354 & -0.9847 & -1.0786 & -0.0939 & 3.2521 & 3.3614 &  0.1093 \\
136   & -51.1212 & -50.9138 & 5.641 & -0.9667 & -0.9887 & -0.0221 & 3.2319 & 3.2689 &  0.0370 \\
\bf{385}   & -51.1388 & -50.9432 & \bf{5.321} & -0.9333 & -0.9413 & \bf{-0.0080} & 3.1883 & 3.2086 &  \bf{0.0204} \\
644   & -51.1510 & -50.9605 & 5.184 & -0.9147 & -0.9125 &  0.0022 & 3.1650 & 3.1700 &  0.0050 \\
\bf{1757}  & -51.1878 & -51.0023 & \bf{5.049} & -0.8563 & -0.8473 &  \bf{0.0090} & 3.0774 & 3.0759 & \bf{-0.0015} \\
2895  & -51.2159 & -51.0267 & 5.149 & -0.8156 & -0.8137 &  0.0019 & 3.0145 & 3.0256 &  0.0111 \\
\bf{3718}  & -51.2306 & -51.0411 & \bf{5.155} & -0.7962 & -0.7934 & \bf{0.0029} & 2.9851 & 2.9903 &  \bf{0.0052} \\
\bf{10470} & -51.2976 & -51.1132 & \bf{5.016} & -0.7087 & -0.7000 &  \bf{0.0087} & 2.8173 & 2.7988 & \bf{-0.0185} \\
\hline
\multicolumn{10}{c}{NO 0.45$-$0.55} \\
51    & -51.0780 & -50.8242 & 6.908 & -1.1764 & -1.2809 & -0.1045 & 3.3410 & 3.4428 &  0.1018 \\
137   & -51.1293 & -50.9092 & 5.989 & -1.0439 & -1.0908 & -0.0468 & 3.2207 & 3.2893 &  0.0686 \\
377   & -51.1588 & -50.9691 & 5.160 & -0.9881 & -0.9696 &  0.0185 & 3.1552 & 3.1447 & -0.0105 \\
649   & -51.1860 & -50.9937 & 5.231 & -0.9383 & -0.9278 &  0.0106 & 3.0893 & 3.0874 & -0.0018 \\
1674  & -51.2464 & --51.051 & 5.305 & -0.8418 & -0.8401 &  0.0017 & 2.9435 & 2.9563 &  0.0128 \\
2854  & -51.2847 & -51.0898 & 5.303 & -0.7859 & -0.7837 &  0.0022 & 2.8513 & 2.8660 &  0.0146 \\
\bf{3572}  & -51.3021 & -51.1065 & \bf{5.324} & -0.7630 & -0.7608 & \bf{0.0022} & 2.8161 & 2.8250 &  \bf{0.0088} \\
\bf{7025}  & -51.3531 & -51.1590 & \bf{5.282} & -0.6953 & -0.6920 & \bf{0.0033} & 2.6850 & 2.6905 &  \bf{0.0055} \\
\bf{10723} & -51.3895 & -51.1937 & \bf{5.326} & -0.6501 & -0.6480 & \bf{0.0021} & 2.5906 & 2.5927 & \bf{0.0020} \\


\hline
\end{tabular}
}
\label{tab:tetrazine-cipsi}
\end{table*}


\clearpage

\subsection{Cyclopentadienone}

\begin{figure}[htb]
\begin{center}
\includegraphics[scale=0.4]{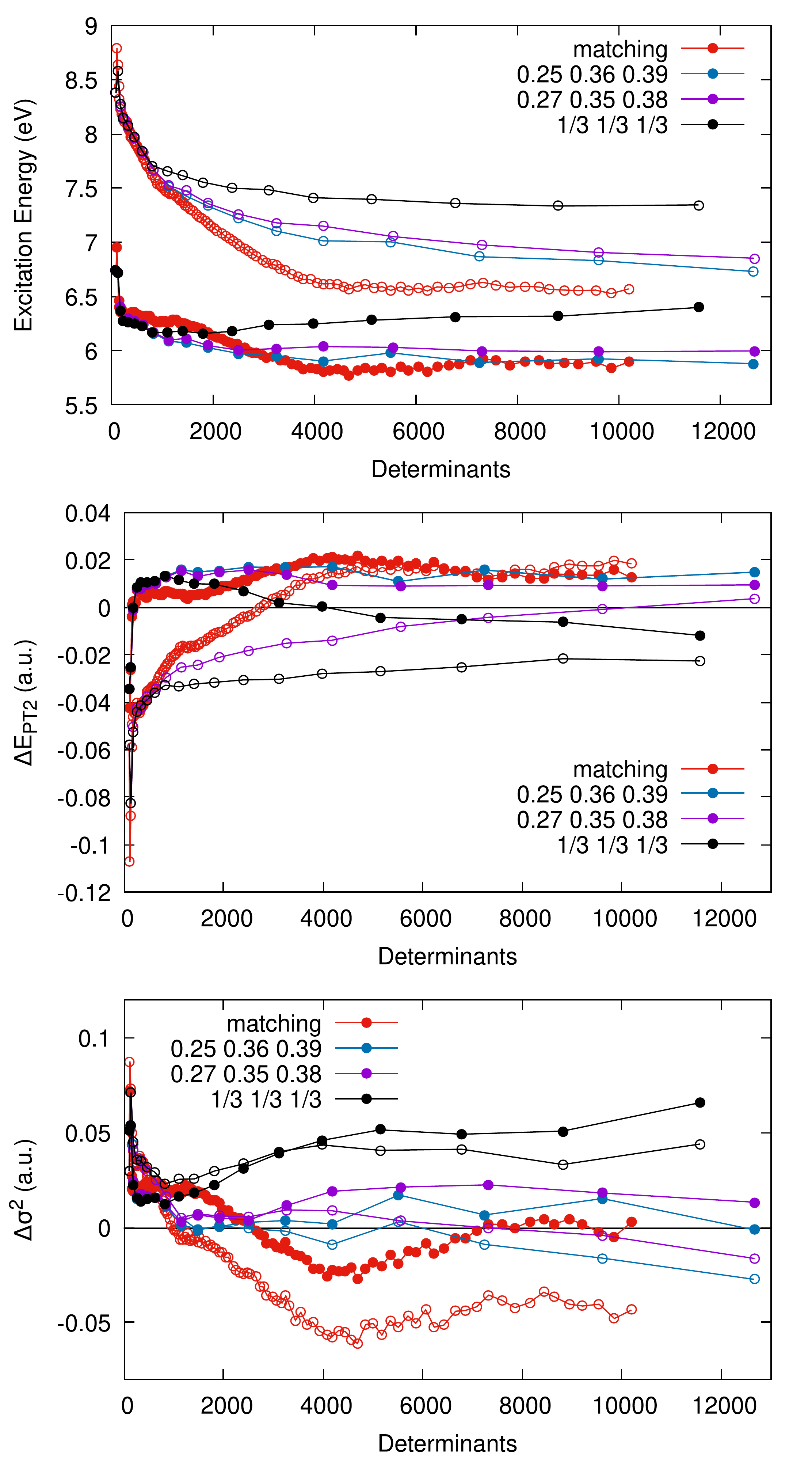}
\caption{\label{fig:cyclopentadienone-cipsi-cas} Cyclopentadienone CIPSI energies, PT2 energies, and variances in the CAS basis, obtained using different input weights ($w_1^{\mathrm{inp}}$ $w_2^{\mathrm{inp}}$ $w_3^{\mathrm{inp}}$) as well as the PT2- and variance-matching algorithm (ii) (labeled `matching'). Filled points represent $\Delta X_{12}$ values while empty points represent $\Delta X_{13}$ values.}
\end{center}
\end{figure}
\begin{figure}[htb]
\begin{center}
\includegraphics[scale=0.4]{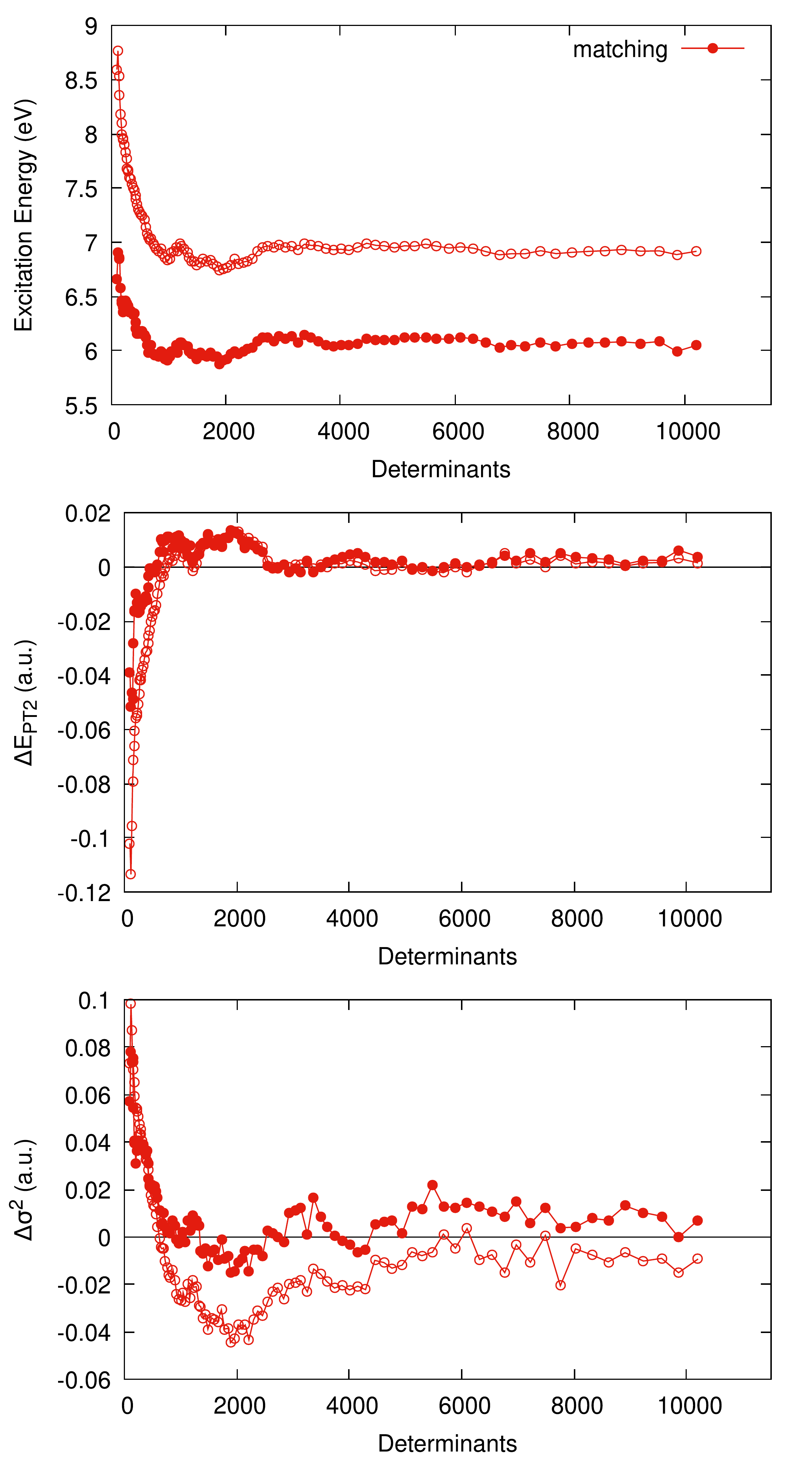}
\caption{\label{fig:cyclopentadienone-cipsi-nos} Cyclopentadienone CIPSI energies, PT2 energies, and variances in the NO basis, obtained using the PT2- and variance-matching algorithm (ii) (labeled `matching'). Filled points represent $\Delta X_{12}$ values while empty points represent $\Delta X_{13}$ values.}
\end{center}
\end{figure}

Data on CAS and NO CIPSI expansions are provided in Table~\ref{tab:cyclopentadienone-cipsi} and Figures~\ref{fig:cyclopentadienone-cipsi-cas} and \ref{fig:cyclopentadienone-cipsi-nos}. Since there are three states to match, we try to achieve a good matching by choosing the input weights manually like for the previous molecules as well as by employing a PT2- and variance-matching algorithm provided in QP (see matching algorithm (ii) in Section~\ref{sec:cipsi-dets-matching}). Using a CAS basis, this algorithm is able to match the PT2 energies of all three states well, but struggles to match the variances of the states simultaneously. Therefore, for the CAS basis, we use input weights to obtain balanced CIPSI expansions. The matching algorithm is instead used for the NO basis since it can simultaneously match all three PT2 energies and variances throughout the expansion.

\clearpage




\begin{table*}[!htb]
\caption{Cyclopentadienone CIPSI energies ($E^{\mathrm{CIPSI}}$), PT2 energies ($E^{\mathrm{PT2}}$), and variances ($\sigma^2$) of ground- and excited-state CIPSI expansions using the matching approach (i) with input weights, $w_I^{\mathrm{inp}}$, and the approach (ii) based on a matching algorithm, with a selection factor of $f=0.02$. The CIPSI excitation energies, $\Delta E$, are given in eV, the remaining values are in a.u. Only the ground-state energies and variances are listed; the values for the other states can be determined from the differences provided, e.g. $\Delta E_{\mathrm{12}}=E(2 ^1\mathrm{A}_1)-E(1 ^1\mathrm{A}_1)$.}
\begin{tabular}{lrrrrrrrrr}
\hline
        & \multicolumn{3}{c}{$E^{\mathrm{CIPSI}}$} & \multicolumn{3}{c}{$E_{\mathrm{PT2}}$} & \multicolumn{3}{c}{$\sigma^2$} \\
\cline{2-4}
\cline{5-7}
\cline{8-10}
\# det. & \multicolumn{1}{c}{$1 ^1\mathrm{A_{1}}$} & \multicolumn{1}{c}{$\Delta E_{12}$} & \multicolumn{1}{c}{$\Delta E_{13}$} & \multicolumn{1}{c}{$1 ^1\mathrm{A_{1}}$} & \multicolumn{1}{c}{$\Delta E_{12}^{\mathrm{PT2}}$} & \multicolumn{1}{c}{$\Delta E_{13}^{\mathrm{PT2}}$} & \multicolumn{1}{c}{$1 ^1\mathrm{A_{1}}$} & \multicolumn{1}{c}{$\Delta\sigma^2_{12}$} & \multicolumn{1}{c}{$\Delta\sigma^2_{13}$}\\
\hline
\multicolumn{10}{c}{CAS 0.27$-$0.35$-$0.38} \\
98    & -45.7035 & 6.742 & 8.375 & -0.9811 & -0.0341 & -0.0577 & 2.9869 &  0.0513 &  0.0299 \\
338   & -45.7525 & 6.294 & 8.067 & -0.8629 &  0.0080 & -0.0426 & 2.8731 &  0.0174 &  0.0357 \\
627   & -45.7566 & 6.237 & 7.827 & -0.8555 &  0.0103 & -0.0343 & 2.8611 &  0.0155 &  0.0252 \\
1904  & -45.7690 & 6.049 & 7.358 & -0.8354 &  0.0148 & -0.0209 & 2.8172 &  0.0052 &  0.0058 \\
2507  & -45.7741 & 6.006 & 7.257 & -0.8277 &  0.0159 & -0.0179 & 2.8031 &  0.0050 &  0.0054 \\
\bf{3227}  & -45.7808 & \bf{6.011} & \bf{7.169} & -0.8183 &  \bf{0.0137} & \bf{-0.0141} & 2.7807 &  \bf{0.0118} &  \bf{0.0065} \\
\bf{7278}  & -45.8074 & \bf{6.006} & \bf{6.981} & -0.7794 &  \bf{0.0097} & \bf{-0.0041} & 2.7007 &  \bf{0.0183} &  \bf{0.0003} \\
\bf{9619}  & -45.8176 & \bf{5.997} & \bf{6.909} & -0.7646 &  \bf{0.0083} & \bf{-0.0020} & 2.6727 &  \bf{0.0199} &  \bf{0.0004} \\
\hline
\multicolumn{10}{c}{CAS } \\
98    & -45.7035 & 6.742 & 8.375 & -0.9811 & -0.0341 & -0.0577 & 2.9869 &  0.0513 &  0.0299 \\
381   & -45.7534 & 6.343 & 8.008 & -0.8616 &  0.0053 & -0.0410 & 2.8710 &  0.0211 &  0.0343 \\
670   & -45.7570 & 6.314 & 7.756 & -0.8549 &  0.0058 & -0.0313 & 2.8597 &  0.0225 &  0.0214 \\
\bf{1019}  & -45.7597 & \bf{6.273} & \bf{7.499} & -0.8503 &  \bf{0.0056} & \bf{-0.0198} & 2.8514 &  \bf{0.0192} & \bf{-0.0018} \\
1729  & -45.7643 & 6.206 & 7.239 & -0.8432 &  0.0068 & -0.0123 & 2.8393 &  0.0145 & -0.0089 \\
2866  & -45.7687 & 5.979 & 6.873 & -0.8363 &  0.0152 &  0.0020 & 2.8298 & -0.0087 & -0.0355 \\
\bf{3036}  & -45.7688 & \bf{5.942} & \bf{6.816} & -0.8362 &  \bf{0.0161} &  \bf{0.0041} & 2.8294 & \bf{-0.0083} & \bf{-0.0363} \\
3551  & -45.7711 & 5.880 & 6.705 & -0.8324 &  0.0185 &  0.0094 & 2.8218 & -0.0147 & -0.0447 \\
5009  & -45.7811 & 5.844 & 6.612 & -0.8180 &  0.0180 &  0.0146 & 2.7895 & -0.0164 & -0.0484 \\
\bf{7248}  & -45.7980 & \bf{5.929} & \bf{6.644} & -0.7926 &  \bf{0.0116} &  \bf{0.0119} & 2.7319 &  \bf{0.0029} & \bf{-0.0321} \\
\bf{10198} & -45.8107 & \bf{5.896} & \bf{6.570} & -0.7745 &  \bf{0.0126} &  \bf{0.0184} & 2.6909 &  \bf{0.0028} & \bf{-0.0434} \\
\hline
\multicolumn{10}{c}{NO } \\
102   & -45.7117 & 6.656 & 8.593 & -1.0544 & -0.0388 & -0.1020 & 2.9870 &  0.0513 &  0.0299 \\
378   & -45.7654 & 6.329 & 7.500 & -0.9294 & -0.0110 & -0.0317 & 2.8498 &  0.0347 &  0.0325 \\
670   & -45.7710 & 5.988 & 7.025 & -0.9183 &  0.0095 & -0.0026 & 2.8338 &  0.0060 & -0.0050 \\
\bf{1025}  & -45.7815 & \bf{5.950} & \bf{6.854} & -0.8984 &  \bf{0.0096} & \bf{0.0069} & 2.807 &  \bf{-0.0006} & \bf{-0.0270} \\
1728  & -45.8040 & 5.979 & 6.836 & -0.8600 &  0.0072 & 0.0079 & 2.7480 & -0.0011 & -0.0307 \\
2834  & -45.8311 & 6.091 & 6.947 & -0.8153 &  0.0010 &  0.0009 & 2.6698 & -0.0021 & -0.0260 \\
\bf{3033}  & -45.8347 & \bf{6.141} & \bf{6.967} & -0.8102 &  \bf{-0.0019} &  \bf{-0.0001} & 2.6524 & \bf{0.0135} & \bf{-0.0137} \\
3488  & -45.8416 & 6.124 & 6.972 & -0.7992 & -0.0001 &  0.0006 & 2.6316 & 0.0087 & -0.0156 \\
4941  & -45.8589 & 6.095 & 6.948 & -0.7727 &  0.0024 &  0.0003 & 2.5790 & 0.0016 & -0.0119 \\
\bf{7151}  & -45.8790 & \bf{6.090} & \bf{6.942} & -0.7415 &  \bf{0.0006} &  \bf{0.0002} & 2.5116 &  \bf{0.0156} & \bf{-0.0079} \\
\bf{10206} & -45.8989 & \bf{6.050} & \bf{6.917} & -0.7133 &  \bf{0.0038} &  \bf{0.0014} & 2.4586 &  \bf{0.0071} & \bf{-0.0092} \\
\hline
\end{tabular}
\label{tab:cyclopentadienone-cipsi}
\end{table*}

\clearpage

\section{More QMC Data}

Tables~\ref{tab:nitroxyl-SI}$-$\ref{tab:cyclopentadienone-SI} show VMC and DMC total energies (a.u.) and excitation energies (eV). The number of determinants (\#~det) and parameters (\#~parm) in each trial wave function is also given. All calculations use BFD pseudopotentials with the corresponding aug-cc-pVDZ basis set.

\clearpage

\subsection{Nitroxyl}

\begin{table*}[!htb]
\caption{Nitroxyl VMC and DMC results. The optimal weights ($w_1^{\mathrm{inp}}-w_2^{\mathrm{inp}}$) in the CAS basis are 0.525$-$0.475. Results for weights of 0.50$-$0.50 are also provided. The optimal weights in the NO basis are 0.50$-$0.50.}
\resizebox{\columnwidth}{!}{
\begin{tabular}{lrrllllll}
\hline
 & & & \multicolumn{3}{c}{VMC} & \multicolumn{3}{c}{DMC} \\
\cline{4-6}
\cline{7-9}
\multicolumn{1}{c}{WF} & \multicolumn{1}{c}{\# det} & \multicolumn{1}{c}{\# parm} & \multicolumn{1}{c}{$E\left(1 ^1\mathrm{A'}\right)$} & \multicolumn{1}{c}{$E\left(2 ^1\mathrm{A'}\right)$} & \multicolumn{1}{c}{$\Delta E$} & \multicolumn{1}{c}{$E\left(1 ^1\mathrm{A'}\right)$} & \multicolumn{1}{c}{$E\left(2 ^1\mathrm{A'}\right)$} & \multicolumn{1}{c}{$\Delta E$} \\
\hline
CAS(2,2)   &    2   &  231 &  -26.4462(3) &  -26.2764(3) &  4.61(1) & -26.4990(2) & -26.3321(2) & 4.54(1) \\
CAS(12,9)  & 3528   & 1598 &  -26.4762(2) &  -26.3134(2) & 4.43(1) & -26.5114(2) & -26.3510(2) & 4.36(1) \\[.5ex]
\hline
\multicolumn{9}{l}{CIPSI (CAS)} \\
0.50$-$0.50  &  363   &  965 & -26.4876(2) & -26.3236(2) & 4.46(1) & -26.5161(2) & -26.3547(2) & 4.39(1) \\
           &  1627  & 2150 & -26.5013(2) & -26.3426(2) & 4.32(1) & -26.5228(2) & -26.3643(2) & 4.31(1) \\
           &  3447  & 2986 & -26.5061(2) & -26.3469(2) & 4.33(1) & -26.5248(2) & -26.3661(2) & 4.32(1) \\
           &  10660 & 5826 & -26.5106(2) & -26.3506(2) & 4.35(1) & -26.5267(2) & -26.3675(2) & 4.33(1) \\[.5ex]
0.525$-$0.475&    322 &  909 & -26.4891(2) & -26.3271(2) & 4.41(1) & -26.5155(2) & -26.3565(2) & 4.33(1) \\
           &   1569 & 2126 & -26.5021(2) & -26.3423(2) & 4.35(1) & -26.5228(2) & -26.3644(2) & 4.31(1) \\
           &   3751 & 3135 & -26.5088(2) & -26.3481(2) & 4.37(1) & -26.5260(2) & -26.3667(2) & 4.33(1) \\
           &   7464 & 4618 & -26.5111(2) & -26.3511(2) & 4.35(1) & -26.5273(2) & -26.3682(2) & 4.33(1) \\
           &  10516 & 5772 & -26.5119(2) & -26.3517(2) & 4.36(1) & -26.5275(2) & -26.3680(2) & 4.34(1) \\[.5ex]
\hline
\multicolumn{9}{l}{CIPSI (NO)} \\
0.50$-$0.50  &    321 &  850 & -26.4917(2) & -26.3322(2) & 4.34(1) & -26.5164(2) &
           -26.3583(2) & 4.30(1) \\
           &   1573 & 1446 & -26.5001(2) & -26.3415(2) & 4.31(1) & -26.5220(2) & -26.3636(2) & 4.31(1) \\
           &   2900 & 2002 & -26.5040(2) & -26.3447(2) & 4.34(1) & -26.5240(2) & -26.3652(2) & 4.32(1) \\
           &   7171 & 4124 & -26.5102(2) & -26.3498(2) & 4.36(1) & -26.5267(2) & -26.3677(2) & 4.33(1) \\
           &  10690 & 5712 & -26.5113(2) & -26.3519(2) & 4.34(1) & -26.5273(2) & -26.3686(2) & 4.32(1) \\
\hline
\multicolumn{8}{l}{TBE$^a$}       & \multicolumn{1}{l}{4.33} \\
\multicolumn{8}{l}{CC3/AVQZ$^b$} & \multicolumn{1}{l}{5.23} \\
\multicolumn{8}{l}{CC4/AVDZ$^c$}               & \multicolumn{1}{l}{4.454} \\
\multicolumn{8}{l}{CC4/AVTZ$^c$}               & \multicolumn{1}{l}{4.380} \\
\multicolumn{8}{l}{CCSDT/AVDZ$^b$}         & \multicolumn{1}{l}{4.756} \\
\multicolumn{8}{l}{CCSDTQ/AVDZ$^b$}        & \multicolumn{1}{l}{4.424} \\
\multicolumn{8}{l}{CCSDTQ/AVTZ$^c$}            & \multicolumn{1}{l}{4.364} \\
\multicolumn{8}{l}{exFCI/AVQZ$^d$}         & \multicolumn{1}{l}{4.32(0)} \\
\multicolumn{8}{l}{CASPT2(12,9)/AVQZ$^d$}  & \multicolumn{1}{l}{4.34} \\
\multicolumn{8}{l}{PC-NEVPT2(12,9)/AVQZ$^d$} & \multicolumn{1}{l}{4.35} \\
\hline
\multicolumn{8}{l}{$^a$Value is `safe', as defined in Ref.~\citenum{questdb}. FCI/AVTZ}\\
\multicolumn{8}{l}{$^b$Ref.~\citenum{loos2021}.}\\
\multicolumn{8}{l}{$^c$Ref.~\citenum{newcc4}.}\\
\multicolumn{8}{l}{$^d$Ref.~\citenum{loos2019}.}\\
\end{tabular}
}
\label{tab:nitroxyl-SI}
\end{table*}


\clearpage

\subsection{Glyoxal}

\begin{table*}[!htb]
\caption{Glyoxal VMC and DMC results. The optimal weights ($w_1^{\mathrm{inp}}-w_2^{\mathrm{inp}}$) in the CAS basis are 0.50$-$0.50 (< 5,000 determinants) and  0.475$-$0.525 (> 5,000 determinants). The optimal weights in the NO basis are 0.45$-$0.55.}
\resizebox{\columnwidth}{!}{
\begin{tabular}{lrrllllll}
\hline
 & & & \multicolumn{3}{c}{VMC} & \multicolumn{3}{c}{DMC} \\ 
\cline{4-6}
\cline{7-9}
\multicolumn{1}{c}{WF} & \multicolumn{1}{c}{\# det} & \multicolumn{1}{c}{\# parm} & \multicolumn{1}{c}{$E\left(1 ^1\mathrm{A_g}\right)$} & \multicolumn{1}{c}{$E\left(2 ^1\mathrm{A_g}\right)$} & \multicolumn{1}{c}{$\Delta E$} & \multicolumn{1}{c}{$E\left(1 ^1\mathrm{A_g}\right)$} & \multicolumn{1}{c}{$E\left(2 ^1\mathrm{A_g}\right)$} & \multicolumn{1}{c}{$\Delta E$} \\
\hline
CAS(8,6)                &    61 &  441 & -44.5603(2) & -44.3473(2) & 5.79(1)   & -44.6428(3)   & -44.4315(3)   & 5.75(1) \\[.5ex]
\hline
\multicolumn{9}{l}{CIPSI (CAS)}\\
0.50$-$0.50                   &   358 &  816 & -44.5729(2) & -44.3633(2) & 5.70(1)   & -44.6437(2)   & -44.4369(2)   & 5.63(1) \\
                        &  1632 & 2683 & -44.5950(2) & -44.3832(2) & 5.76(1)   & -44.6514(3)   & -44.4434(3)   & 5.66(1) \\
                        &  3446 & 3903 & -44.6075(2) & -44.3965(2) & 5.74(1)   & -44.6565(2)   & -44.4494(2)   & 5.64(1) \\
                        & 10342 & 7248 & -44.6273(2) & -44.4124(2) & 5.85(1)   & -44.6663(2)   & -44.4567(2)   & 5.70(1) \\[.5ex]
0.475$-$0.525           & 10660 & 7374 & -44.6288(2) & -44.4160(2) & 5.79(1) & -44.6675(2) & -44.4585(2) & 5.69(1) \\[.5ex]
\hline
\multicolumn{9}{l}{CIPSI (NO)}\\
0.45$-$0.55   &  3749 & 2908 & -44.6072(2) & -44.3957(2) & 5.76(1)   & -44.6564(2)   & -44.4482(2)   & 5.67(1) \\
                        &  5046 & 3498 & -44.6127(2) & -44.4004(2) & 5.78(1)   & -44.6600(2)   & -44.4509(2)   & 5.69(1) \\
                        &  7040 & 4500 & -44.6150(2) & -44.4048(2) & 5.72(1)   & -44.6613(2)   & -44.4534(2)   & 5.66(1) \\
                        &  8642 & 5433 & -44.6181(2) & -44.4098(2) & 5.67(1)   & -44.6632(2)   & -44.4559(2)   & 5.64(1) \\
                        & 10010 & 6188 & -44.6205(2) & -44.4114(2) & 5.69(1)   & -44.6635(2)   & -44.4567(2)   & 5.63(1) \\
                        & 12132 & 6269 & -44.6237(2) & -44.4145(2) & 5.70(1)   & -44.6654(2)   & -44.4576(2)   & 5.66(1) \\
                        & 15349 & 8577 & -44.6276(2) & -44.4180(2) & 5.70(1)   & -44.6670(2)   & -44.4600(2)   & 5.63(1) \\
\hline
\multicolumn{8}{l}{TBE$^a$}         &  \multicolumn{1}{l}{5.61} \\
\multicolumn{8}{l}{CC3/AVQZ$^b$}            &  \multicolumn{1}{l}{6.76} \\
\multicolumn{8}{l}{CC4/6-31+G*$^c$}           &  \multicolumn{1}{l}{5.699} \\
\multicolumn{8}{l}{CC4/AVDZ$^c$}              &  \multicolumn{1}{l}{5.593} \\
\multicolumn{8}{l}{CCSDTQ/6-31+G*$^c$}        &  \multicolumn{1}{l}{5.670} \\
\multicolumn{8}{l}{exFCI/AVDZ}    &  \multicolumn{1}{l}{5.56(11)} \\
\multicolumn{8}{l}{SAC-CI/AVDZ$^d$}    &  \multicolumn{1}{l}{5.66} \\
\multicolumn{8}{l}{CASPT2(14,12)/AVQZ$^b$}    &  \multicolumn{1}{l}{5.43} \\
\multicolumn{8}{l}{PC-NEVPT2(14,12)/AVQZ$^b$} &  \multicolumn{1}{l}{5.52} \\
\hline
\multicolumn{8}{l}{$^a$This value is `safe', as defined in Ref.~\citenum{questdb}. exFCI/AVDZ + ( CCSDT/AVTZ - CCSDT/AVDZ ) } \\
\multicolumn{8}{l}{$^b$Ref.~\citenum{loos2019}.} \\
\multicolumn{8}{l}{$^c$Ref.~\citenum{newcc4}.} \\
\multicolumn{8}{l}{$^d$Ref.~\citenum{SACCI}. See reference for basis set details.} \\
\end{tabular}
}
\label{tab:glyoxal-SI}
\end{table*}


\clearpage

\subsection{Tetrazine}

\begin{table*}[!htb]
\caption{Tetrazine VMC and DMC results. The optimal weights ($w_1^{\mathrm{inp}}-w_2^{\mathrm{inp}}$) in the  CAS and NO basis are 0.45$-$0.55.}
\resizebox{\columnwidth}{!}{
\begin{tabular}{lrrllllll}
\hline
 & & & \multicolumn{3}{c}{VMC} & \multicolumn{3}{c}{DMC} \\ 
\cline{4-6}
\cline{7-9}
\multicolumn{1}{c}{WF} & \multicolumn{1}{c}{\# det} & \multicolumn{1}{c}{\# parm} & \multicolumn{1}{c}{$E\left(1 ^1\mathrm{A_{1g}}\right)$} & \multicolumn{1}{c}{$E\left(2 ^1\mathrm{A_{1g}}\right)$} & \multicolumn{1}{c}{$\Delta E$} & \multicolumn{1}{c}{$E\left(1 ^1\mathrm{A_{1g}}\right)$} & \multicolumn{1}{c}{$E\left(2 ^1\mathrm{A_{1g}}\right)$} & \multicolumn{1}{c}{$\Delta E$} \\
\hline
CAS(4,4)             &     6 &  399 & -52.1492(2) & -51.9510(2) & 5.39(1)  & -52.2647(3)   & -52.0734(3)   & 5.21(1) \\
CAS(8,6)             &    33 &  414 & -52.1649(2) & -51.9569(2) & 5.66(1)  & -52.2725(3)   & -52.0767(3)   & 5.33(1) \\
CAS(12,8)            &   112 &  454 & -52.1652(2) & -51.9658(2) & 5.43(1)  & -52.2740(3)   & -52.0824(3)   & 5.21(1)  \\
CAS(14,10)           &  1824 & 1078 & -52.1772(2) & -51.9880(2) & 5.15(1)  & -52.2804(3)   & -52.0927(3)   & 5.11(1) \\[.5ex]
\hline
\multicolumn{9}{l}{CIPSI (CAS)} \\
0.45$-$0.55   &   385 &  871 & -52.1826(2) & -52.0008(2) & 4.95(1)  & -52.2811(3)   & -52.0985(3)   & 4.97(1) \\
                     &  1757 & 1608 & -52.1966(2) & -52.0203(2) & 4.80(1)  & -52.2841(2)   & -52.1042(2)   & 4.89(1) \\
                     &  3718 & 2302 & -52.2138(2) & -52.0323(2) & 4.94(1)  & -52.2918(3)   & -52.1089(3)   & 4.98(1) \\
                     & 10470 & 6934 & -52.2321(2) & -52.0504(2) & 4.94(1)  & -52.2996(2)   & -52.1171(2)   & 4.97(1) \\[.5ex]
\hline
\multicolumn{9}{l}{CIPSI (NO)} \\
0.45$-$0.55   &  3572 & 2414 & -52.2198(2) & -52.0346(2) & 5.04(1)  & -52.2935(3)   & -52.1098(3)   & 5.00(1) \\
                     &  7025 & 4141 & -52.2300(2) & -52.0451(2) & 5.03(1)  & -52.2981(3)   & -52.1148(3)   & 4.99(1) \\
                     & 10723 & 6053 & -52.2379(2) & -52.0528(2) & 5.04(1)  & -52.3015(3)   & -52.1182(3)   & 4.99(1) \\
\hline
\multicolumn{8}{l}{TBE$^a$}                  &  \multicolumn{1}{l}{4.61} \\
\multicolumn{8}{l}{CC3/6-31+G*$^b$}                 &  \multicolumn{1}{l}{6.22} \\
\multicolumn{8}{l}{CC3/AVQZ$^b$}                 &  \multicolumn{1}{l}{6.19} \\
\multicolumn{8}{l}{CC4/6-31+G*$^c$}              &  \multicolumn{1}{l}{5.06} \\
\multicolumn{8}{l}{CCSDT/AVTZ$^b$}               &  \multicolumn{1}{l}{5.96} \\
\multicolumn{8}{l}{exFCI/AVDZ}               &  \multicolumn{1}{l}{5.15(3)} \\
\multicolumn{8}{l}{CASPT2(14,10)/AVQZ$^b$}       &  \multicolumn{1}{l}{4.68} \\
\multicolumn{8}{l}{PC-NEVPT2(14,10)/AVQZ$^b$}    &  \multicolumn{1}{l}{4.60} \\
\hline
\multicolumn{8}{l}{$^a$This value is `unsafe', as defined in Ref~\citenum{questdb}. NEVTPT2/AVTZ} \\
\multicolumn{8}{l}{$^b$Ref.~\citenum{loos2019}.} \\
\multicolumn{8}{l}{$^c$Ref.~\citenum{newcc4}.} \\
\end{tabular}
}
\label{tab:tetrazine}
\end{table*}



\subsection{Cyclopentadienone}

\clearpage

\begin{sidewaystable}[!htb]
\caption{Cyclopentadienone VMC and DMC results. The optimal weights ($w_1^{\mathrm{inp}}-w_2^{\mathrm{inp}}-w_3^{\mathrm{inp}}$) in the CAS basis are 0.27$-$0.35$-$0.38.}
\resizebox{\columnwidth}{!}{
\begin{tabular}{lrrllllllllllll}
\hline
 & & & \multicolumn{6}{c}{VMC} & \multicolumn{6}{c}{DMC} \\ 
\cline{4-9}
\cline{10-15}
\multicolumn{1}{c}{WF} & \multicolumn{1}{c}{\# det} & \multicolumn{1}{c}{\# parm} & \multicolumn{1}{c}{$E\left(1 ^1\mathrm{A_{1}}\right)$} & \multicolumn{1}{c}{$E\left(2 ^1\mathrm{A_{1}}\right)$} & \multicolumn{1}{c}{$E\left(3 ^1\mathrm{A_{1}}\right)$} & \multicolumn{1}{c}{$\Delta E_{12}$} & \multicolumn{1}{c}{$\Delta E_{13}$} & \multicolumn{1}{c}{$\Delta E_{23}$} & \multicolumn{1}{c}{$E\left(1 ^1\mathrm{A_{1}}\right)$} & \multicolumn{1}{c}{$E\left(2 ^1\mathrm{A_{1}}\right)$} & \multicolumn{1}{c}{$E\left(3 ^1\mathrm{A_{1}}\right)$} & \multicolumn{1}{c}{$\Delta E_{12}$} & \multicolumn{1}{c}{$\Delta E_{13}$} & \multicolumn{1}{c}{$\Delta E_{23}$} \\
\hline
CAS(6,6)            &  208 &  967 & -46.7358(1) & -46.5119(1) & -46.4438(1) & 6.093(4) & 7.947(4) & 1.854(4) & -46.8322(1) & -46.6109(1) & -46.5439(1) & 6.021(6) & 7.844(6) & 1.822(6) \\[.5ex]
\hline
\multicolumn{15}{l}{CIPSI (CAS)} \\
0.27$-$0.35$-$0.38   &  9619 & 10062 & -46.7794(2) & -46.5591(2) & -46.5220(2)  & 6.00(1) & 7.00(1) & 1.01(1) & -46.8488(2) & -46.6329(2) & -46.5967(2) & 5.87(1) & 6.86(1) & 0.99(1) \\[0.5ex]
matching ($f$: 0.02) &  1019 &  4735 & -46.7532(2) & -46.5281(2) & -46.4930(2) & 6.13(1) & 7.08(1) & 0.96(1) & -46.8405(2) & -46.6204(2) & -46.5854(2) & 5.99(1) & 6.94(1) & 0.95(1) \\
                     &  3036 &  6235 & -46.7663(2) & -46.5440(2) & -46.5099(2) & 6.05(1) & 6.98(1) & 0.93(1) & -46.8448(2) & -46.6278(2) & -46.5938(2) & 5.91(1) & 6.83(1) & 0.92(1) \\
                     &  7248 &  9008 & -46.7741(2) & -46.5537(2) & -46.5200(2) & 6.00(1)  & 6.91(1) & 0.92(1) & -46.8472(2) & -46.6310(2) & -46.5972(2) & 5.88(1) & 6.80(1) & 0.92(1) \\
                     & 10198 & 10557 & -46.7807(2) & -46.5605(2) & -46.5263(2) & 5.99(1)  & 6.92(1) & 0.93(1) & -46.8503(2) & -46.6330(2) & -46.5988(2) & 5.90(1) & 6.84(1) & 0.99(1) \\
\hline
\multicolumn{15}{l}{CIPSI (NO)} \\
matching ($f$: 0.02)  &  1025 &  2403 & -46.7542(2) & -46.5361(2) & -46.5000(2) & 5.94(1) & 6.92(1) & 0.98(1) & -46.8389(2) & -46.6222(2) & -46.5864(2) & 5.90(1) & 6.87(1) & 0.97(1) \\
                    &  3033 &  3645 & -46.7670(2) & -46.5485(2) & -46.5112(2) & 5.95(1) & 6.96(1) & 1.01(1) & -46.8432(2) & -46.6276(2) & -46.5908(2) & 5.87(1) & 6.87(1) & 1.00(1) \\
                    &  7151 &  5365 & -46.7826(2) & -46.5610(2) & -46.5236(2) & 6.03(1)  & 7.05(1) & 1.02(1) & -46.8480(2) & -46.6302(2) & -46.5937(2) & 5.92(1) & 6.91(1) & 0.99(1) \\
                    & 10206 &  6574 & -46.7874(2) & -46.5656(2) & -46.5292(2) & 6.04(1)  & 7.03(1) & 0.99(1) & -46.8498(2) & -46.6330(2) & -46.5965(2) & 5.90(1) & 6.89(1) & 0.99(1) \\
\hline
\multicolumn{12}{l}{TBE$^a$} & \multicolumn{1}{l}{6.00$^b$} & \multicolumn{1}{l}{6.09$^c$} & \multicolumn{1}{l}{0.09} \\
\multicolumn{12}{l}{ADC(3)/AVTZ}  & \multicolumn{1}{l}{4.59$^d$} & \multicolumn{1}{l}{6.50$^d$} & \multicolumn{1}{l}{1.91} \\
\multicolumn{12}{l}{CC2/AVTZ}     &              & \multicolumn{1}{l}{6.50$^d$} & \multicolumn{1}{l}{} \\
\multicolumn{12}{l}{CC3/AVTZ}     & \multicolumn{1}{l}{7.10$^d$} & \multicolumn{1}{l}{6.21$^d$} & \multicolumn{1}{l}{-0.89} \\
\multicolumn{12}{l}{CCSD/AVTZ}    &              & \multicolumn{1}{l}{6.68$^d$} & \multicolumn{1}{l}{} \\
\multicolumn{12}{l}{CCSDT-3/AVTZ} &              & \multicolumn{1}{l}{6.33$^d$} & \multicolumn{1}{l}{} \\
\multicolumn{12}{l}{exFCI/AVDZ}   & \multicolumn{1}{l}{5.81(39)} & \multicolumn{1}{l}{6.93(24)} & \multicolumn{1}{l}{1.10(13)} \\
\hline
\multicolumn{15}{l}{$^a$This value is `unsafe' as defined in Ref.~\citenum{questdb}.} \\
\multicolumn{15}{l}{$^b$NEVPT2/AVTZ.} \\
\multicolumn{15}{l}{$^c$CCSDT/AVDZ + ( CC3/AVTZ-CC3/AVDZ ).} \\
\multicolumn{15}{l}{$^d$Values are also `unsafe'.} \\
\end{tabular}
}
\label{tab:cyclopentadienone-SI}
\end{sidewaystable}

\clearpage

\section{State-specific VMC optimization}\label{sec:algo-dets}

\subsection{Gradient Derivation}

The state-specific optimization method used here minimizes the objective function
\begin{equation}\label{eq:objfunc}
O_I[\Psi_I]=E_I[\Psi_I] + \sum_{J\ne I}\lambda_{IJ}|S_{IJ}|^2,
\end{equation}
where $E_I$ is given by,
\begin{equation}\label{eq:energy}
E_I=\frac{\langle \Psi_I | H | \Psi_I \rangle}{\langle \Psi_I | \Psi_I \rangle}=\langle  E_{\mathrm{L}}^I \left(\mathbf{R}\right) \rangle_{\mathbf{R}\sim |\Psi_I \left(\mathbf{R}\right)|^2},
\end{equation}

For a VMC calculation over multiple states, a guiding function is introduced, $\Psi_{\mathrm{g}}=\sqrt{\sum_{I}|\Psi_I|^2}$ and one computes the energies of the states as
\begin{equation}\label{eq:eloc}
E_I=\frac{\left\langle t_I E_{\mathrm{L}}^I \right\rangle_{\mathbf{R}\sim\rho_{\mathrm{g}}(\mathbf{R})}}{\left\langle t_I \right\rangle_{\mathbf{R}\sim\rho_{\mathrm{g}}(\mathbf{R})}},
\end{equation}
where $E_{\mathrm{L}}^I$ is the local energy of state $I$ and $t_I=|\Psi_I/\Psi_{\mathrm{g}}|^2$. 

Taking the derivative of Eq~\ref{eq:objfunc} with respect to a parameter of state $I$, $p_{I}$, gives
\begin{eqnarray}\label{eq:obder}
\frac{\partial O_I}{\partial p_{I}}&=&2\left\{ \left\langle t_I E_{\mathrm{L}}^I \frac{\partial}{\partial p_{I}}\ln{\Psi_I} \right\rangle \left\langle t_I \right\rangle^{-1} - \left\langle t_I E_{\mathrm{L}}^I \right\rangle \left\langle t_I \frac{\partial}{\partial p_{I}}\ln{\Psi_I} \right\rangle \left\langle t_I \right\rangle^{-2} \right\} \\ 
&+& 2\sum_{J\ne I}\lambda_{IJ} \left\{ \left\langle \frac{\Psi_I}{\Psi_{\mathrm{g}}}\frac{\Psi_J}{\Psi_{\mathrm{g}}}\frac{\partial}{\partial p_{I}}\ln{\Psi_I}\right\rangle \frac{\left\langle \frac{\Psi_I}{\Psi_{\mathrm{g}}}\frac{\Psi_J}{\Psi_{\mathrm{g}}} \right\rangle}{\left\langle t_I \right\rangle \left\langle t_J \right\rangle} - \left\langle t_I \frac{\partial}{\partial p_{I}}\ln{\Psi_I} \right\rangle \frac{\left\langle \frac{\Psi_I}{\Psi_{\mathrm{g}}}\frac{\Psi_J}{\Psi_{\mathrm{g}}} \right\rangle^2}{\left\langle t_I \right\rangle \left\langle t_J \right\rangle} \left\langle t_I \right\rangle^{-1} \right\},\nonumber
\end{eqnarray}
where $\left\langle \cdot \right\rangle$ implies the average over configurations sampled from $\rho_{\mathrm{g}}(\mathbf{R})$. 
Therefore, the gradient has two terms
\begin{equation}
\frac{\partial O_I}{\partial p_{I}}=\left[g_{\mathrm{tot}}\right]_{I}=
\left[g_{\mathrm{E}} + g_{\mathrm{O}}\right]_{I},
\end{equation}
where the first term, $g_{\mathrm{E}}$, is the derivative of the energy (Eq.~\ref{eq:eloc}) while the second term, $g_{\mathrm{O}}$, is the derivative of the orthogonality penalty (the second term in Eq.~\ref{eq:objfunc}). 
The term $g_{\mathrm{O}}$ can be rewritten as
\begin{equation}
 \left[g_{\mathrm{O}}\right]_{I}=2\sum_{J\ne I}\lambda_{IJ} |S_{IJ}|^2 \left\{ \left\langle \frac{\Psi_I}{\Psi_{\mathrm{g}}} \frac{\Psi_J}{\Psi_{\mathrm{g}}} \frac{\partial}{\partial p_I}\ln{\Psi_I}\right\rangle \left\langle \frac{\Psi_I}{\Psi_{\mathrm{g}}} \frac{\Psi_J}{\Psi_{\mathrm{g}}} \right\rangle^{-1} - \left\langle t_I \frac{\partial}{\partial p_I}\ln{\Psi_I} \right\rangle \left\langle t_I \right\rangle^{-1} \right\},
\end{equation}
using that the squared overlap is given by
\begin{equation}\label{eq:overlap}
 |S_{IJ}|^2=\frac{ \left\langle \frac{\Psi_I}{\Psi_{\mathrm{g}}} \frac{\Psi_J}{\Psi_{\mathrm{g}}} \right\rangle^2 }{ \left\langle \frac{\Psi_I^2}{\Psi^2_{\mathrm{g}}} \right\rangle \left\langle \frac{\Psi_J^2}{\Psi^2_{\mathrm{g}}} \right\rangle }=\frac{\left\langle \frac{\Psi_I}{\Psi_{\mathrm{g}}} \frac{\Psi_J}{\Psi_{\mathrm{g}}} \right\rangle^2}{\left\langle t_I \right\rangle \left\langle t_J \right\rangle}.
\end{equation}
For each state $I$, the total gradient is then used in the stochastic reconfiguration (SR) equations,
\begin{equation}\label{eq:sr}
\bar{S}^{(I)}\Delta p_I = -\tau [g_{\mathrm{tot}}]_I,
\end{equation}
where $\bar{S}^{(I)}$ is related to the overlap matrix between the derivatives of the wave function of state $I$ with respect to its parameters.

\clearpage

\subsection{Optimization of 2-determinant wave function of nitroxyl}\label{sec:algo-dets-lambda}

\begin{figure}[htb]
\begin{center}
\includegraphics[width=\columnwidth]{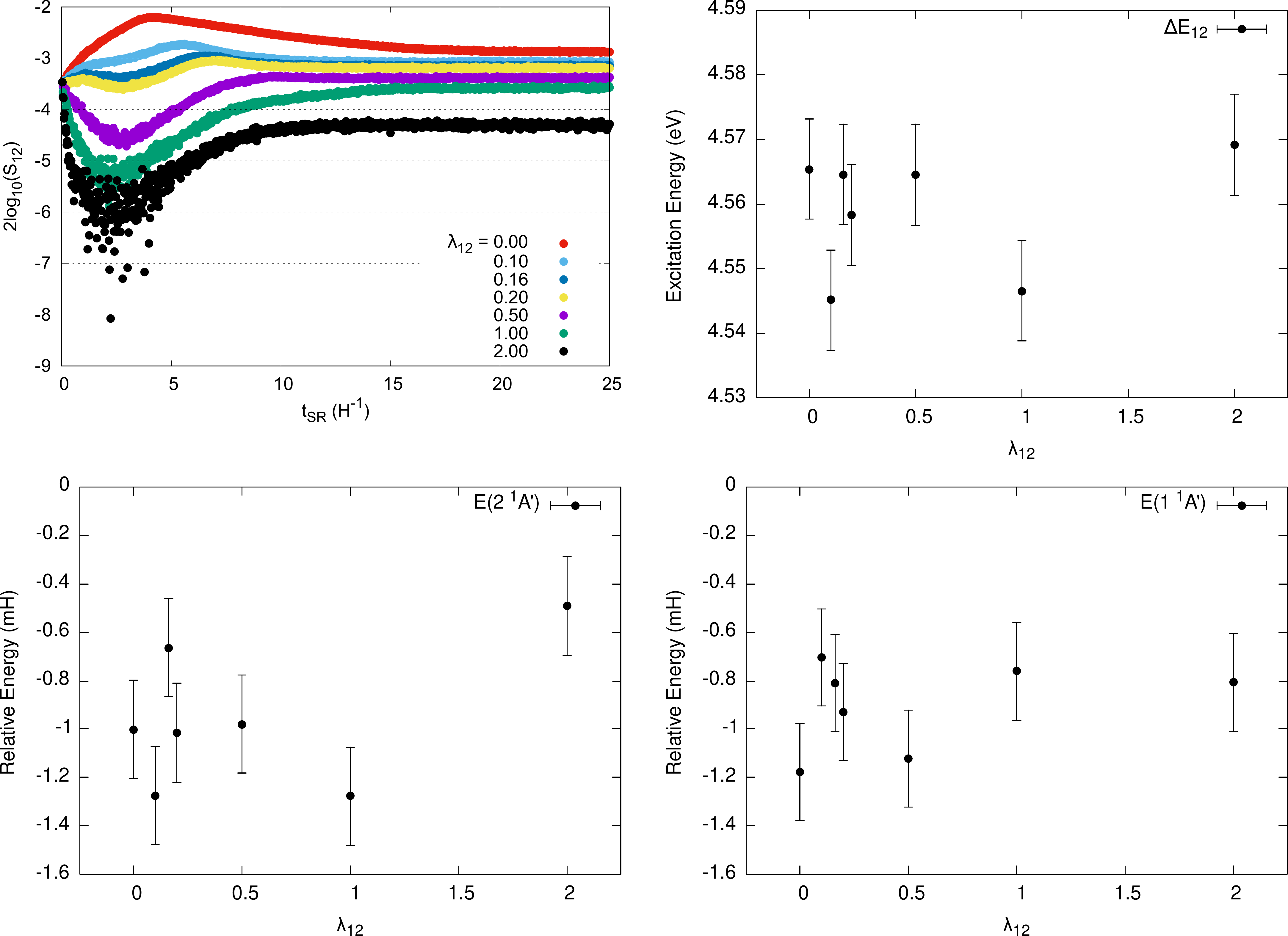}
\caption{\label{fig:nitroxyl-overlap} Effect of choice of $\lambda_{12}$ on the VMC optimization of a 2-determinant wave function for the ground and excited states of nitroxyl. The starting wave function comprises the HF and the relevant doubly-excited determinant, expressed on the CAS orbitals. Top-left: The squared overlap of the ground and excited states ($|S_{12}|^2$) plotted along the SR optimization for different values of $\lambda_{12}$ (a SR time-step of 0.025 H$^{-1}$ is used). Top-right: Final VMC excitation energies obtained with different $\lambda_{12}$. Bottom: Final VMC energies of the ground (left) and excited state (right) shifted by 26.4460~H and 26.2784~H, respectively, using different $\lambda_{12}$.}
\end{center}
\end{figure}



The 2-determinant wave function of nitroxyl studied in Fig.~\ref{fig:nitroxyl-overlap} shows that, without any penalty on the energy ($\lambda_{12}=0$), the excited state does not collapse to the ground state. This is due to the orbital symmetries in the dominant determinant of the double excitation. The doubly-excited determinant promotes both electrons to an orbital of different symmetry than the original ($\mathrm{a'}\rightarrow\mathrm{a''}$) and orbital optimization cannot transform the doubly-excited determinant into the HF one. In this case, $|S_{IJ}|^2$ always plateaus, and increasing $\lambda_{12}$ reduces the value at which it plateaus. All values of the excitation energies (top-right of Fig.~\ref{fig:nitroxyl-overlap}) are compatible within the error bars.





\clearpage

\bibliography{bibliography}